\documentclass[aps,preprint,groupedaddress,showpacs,floatfix]{revtex4}
\usepackage{graphicx}

\begin{document}

\title{Comparative study of the scalar- and tensor-meson production in the reaction $\gamma\gamma^*(Q^2)\to \eta\pi^0$}
\author {
N.N. Achasov$^{\,a}$ \email{achasov@math.nsc.ru}, A.V.
Kiselev$^{\,a,b}$, \email{kiselev@math.nsc.ru} and G.N. Shestakov$^{\,a}$ \email{shestako@math.nsc.ru}}

\affiliation{
   $^a$Laboratory of Theoretical Physics,
 Sobolev Institute for Mathematics, 630090, Novosibirsk, Russia\\
$^b$Novosibirsk State University, 630090, Novosibirsk, Russia}

\date{\today}

\begin{abstract}

The prediction of the cross section $\sigma(\gamma\gamma^*(Q^2)\to
\eta\pi^0)$ based on the simultaneous description of the Belle
data on the $\gamma\gamma\to \eta\pi^0$ reaction and the KLOE data
on the $\phi\to\eta\pi^0\gamma$ decay is presented. The production
of the scalar $a_0(980)$ and tensor $a_2(1320)$ is studied in
detail. It is shown that the QCD based asymptotics of the
$\gamma^*(Q^2)\gamma\to a_2(1320)\to\eta\pi^0$ cross section can
be reached by the compensation of the contributions of $\rho(770)$
and $\omega(782)$ with the contributions of their radial
excitations in $Q^2$ channel. At large $Q^2$ the $a_2(1320)$
contribution is expected to be dominant, while at $Q^2=0$ it is
similar to the scalars contribution.

\end{abstract}
 \maketitle

\section{Introduction}

Study of the nature of light scalars $a_0(980)$ and $f_0(980)$,
well-established part of the proposed light scalar meson nonet
\cite{pdg-2014}, is one of the central problems of nonperturbative
QCD, it is important for understanding the way chiral symmetry is
realized in the low energy region and, consequently, for the
understanding of confinement. Many papers have already been
devoted to this subject, see, for example, Refs.
\cite{closeIK,tornquist,oller-2012,itep,rosner,klempt,volkovRadzhabov,harada}.

Naively one could think that the scalar $a_0(980)$ and $f_0(980)$
mesons are the $q\bar q$ P-wave states with the same quark
structure as $a_2(1320)$ and $f_2(1270)$, respectively. But now
there are many indications that the above scalars are four-quark
states, see, for example, Refs.
\cite{jaffe,joe,pennington,fourQuarkGG,AS88,achasov-89,achasov-97,SNDa0,f0exp,a0f0,kloea0,our_a0,achasov-03,achasov-08,annsgn2010,annsgn2011}
and references therein.

One of these indications is the suppression of the $a_0(980)$ and
$f_0(980)$ resonances in the $\gamma\gamma\to\eta\pi^0$ and
$\gamma\gamma\to\pi\pi$ reactions, respectively, predicted in 1982
\cite{fourQuarkGG} and confirmed by experiment \cite{pdg-2014},
see \cite{GGwidths}. The elucidation of the mechanisms of the
$\sigma(600)$, $f_0(980)$, and $a_0(980)$ resonance production in
the $\gamma\gamma$ collisions confirmed their four-quark structure
\cite{AS88,achasov-08,annsgn2010,annsgn2011}. Light scalar mesons
are produced in $\gamma\gamma$ collisions mainly via
rescatterings, that is, via the four-quark transitions. As for the
$a_2(1320)$ and $f_2(1270)$, the well-known $q\bar q$ states, they
are produced mainly via the two quark transitions (direct
couplings with $\gamma\gamma$).

Another argument in favor of the four-quark nature of the
$a_0(980)$ and $f_0(980)$ is the fact that the $\phi(1020)\to
a_0\gamma$ and $\phi(1020)\to f_0\gamma$ decays proceed through
the kaon loop: $\phi\to K^+K^-\to a_0\gamma$, $\phi\to K^+K^-\to
f_0\gamma$, i.e. via the four-quark transition
\cite{achasov-89,achasov-97,a0f0,our_a0,achasov-03}. The kaon loop
model was suggested in Ref. \cite{achasov-89} and confirmed by the
experiment ten years later \cite{SNDa0,f0exp,kloea0}.

Recently the comparative study of the scalar and tensor mesons was
proposed in the $e^+e^-\to \gamma^*\to (a_0+a_2)\gamma\to
\eta\pi^0\gamma$ and $e^+e^-\to \gamma^*\to (f_0+f_2)\gamma\to
\pi^0\pi^0\gamma$ reactions (i.e. in the timelike region of
$\gamma^*$) \cite{agkr-2013}.

In the present study we consider the reaction $\gamma^*
\gamma\to\pi^0\eta$ in the spacelike region of $\gamma^*$.

In 2009 Belle Collaboration published high-statistical data on the
$\gamma\gamma\to \eta\pi^0$ reaction \cite{uehara}. These data
revealed the specific feature of the $\gamma\gamma\to\pi^0\eta$
cross section: it turned out sizable in the region between the
$a_0(980)$ and $a_2 (1320)$ resonances, that certainly indicates
the presence of additional contributions. The experimenters took
into account the putative heavy isovector scalar $a_0'$ with mass
about $1.3$ GeV (they called it $a(Y)$) along with $a_0(980)$ and
$a_2(1320)$ together with the polynomial coherent background
\cite{uehara}.

In the theoretical works Refs. \cite{annsgn2010,annsgn2011}
$a_0(980)$ is produced mainly by loops, and the Born contribution
to $\gamma\gamma\to\eta\pi^0$ plays role of coherent background,
see Figs. \ref{Amplitudes}, \ref{BornVAKK}. The kaon formfactor
$G_{K^+}(t,u)$ in loop diagrams $\gamma\gamma\to K^+K^-\to a_0$
was introduced in those papers.

The given paper is the development of the Refs.
\cite{annsgn2010,annsgn2011}. Basing on their theoretical results,
we perform new fitting of the Belle data simultaneously with the
KLOE data on the $\phi\to\eta\pi^0\gamma$ decay. We show that in
the kaon loop model of the $\gamma\gamma\to K^+ K^-\to a_0$
transition the kaon formfactor $G_{K^+}(t,u)$ is not required for
good data description, as well as in the $\phi\to K^+K^-\to
a_0\gamma\to\eta\pi^0\gamma$ decay and $\phi\to K^+K^-\to
f_0\gamma\to\pi^0\pi^0\gamma$
\cite{achasov-89,achasov-97,SNDa0,f0exp,a0f0,achasov-03,kloea0,our_a0}.
The obtained $a_0(980)$ coupling constants are in good agreement
with the four-quark model prediction \cite{achasov-89}.

Using results on $\gamma\gamma\to \eta\pi^0$ we predict the
$\sigma(\gamma^*(Q^2)\gamma\to \eta\pi^0,s)$, where $s$ is the
$\gamma^*\gamma$ invariant mass and $Q^2$ is the $\gamma^*$
virtuality.

The measurement of the cross section
$\sigma(\gamma^*(Q^2)\gamma\to \eta\pi^0,s)$ would allow
additional check of the models of the $a_0(980)$ structure and our
understanding of the mechanism of the reactions $\gamma^*\gamma\to
\eta\pi^0$ as well as $\gamma\gamma\to \eta\pi^0$.

The theoretical description of the $\gamma^*\gamma\to \eta\pi^0$
reaction is in Sec. \ref{formulas}. The results on the
$\gamma\gamma\to \eta\pi^0$ data description are presented in Sec.
\ref{results}. The prediction of the
$\sigma(\gamma^*(Q^2)\gamma\to \eta\pi^0,s)$ is in Sec.
\ref{vectorExcitations}. It is found that the role of vector
excitations ($\rho'$, $\omega'$, etc.) is crucial for the
$a_2(1320)$ and $f_2(1270)$ production: the QCD based asymptotics
of the $\gamma^*(Q^2)\gamma\to a_2(1320)\to\eta\pi^0$ (or
$\gamma^*(Q^2)\gamma\to f_2(1270)\to\pi^0\pi^0$) cross section can
be reached only by taking into account cancellation of the
contributions of $\rho(770),\omega(782)$ and $\phi(1020)$ in $Q^2$
channel with the contributions of their radial excitations. The
$a_2(1320)$ contribution is expected to be dominant at large
$Q^2$, while at $Q^2=0$ it is similar to the $a_0(980)+a_0'$
contribution. The conclusion is in Sec. \ref{conclusion}. Some
details are provided in Appendices I-III.

Noticed misprints of Refs. \cite{annsgn2010,annsgn2011} are
mentioned in Ref. \cite{misprints}.

\section{Theoretical description of the $\gamma^*\gamma\to \eta\pi^0$ reaction}
\label{formulas}

\begin{figure} \centerline{\includegraphics[width=15cm]{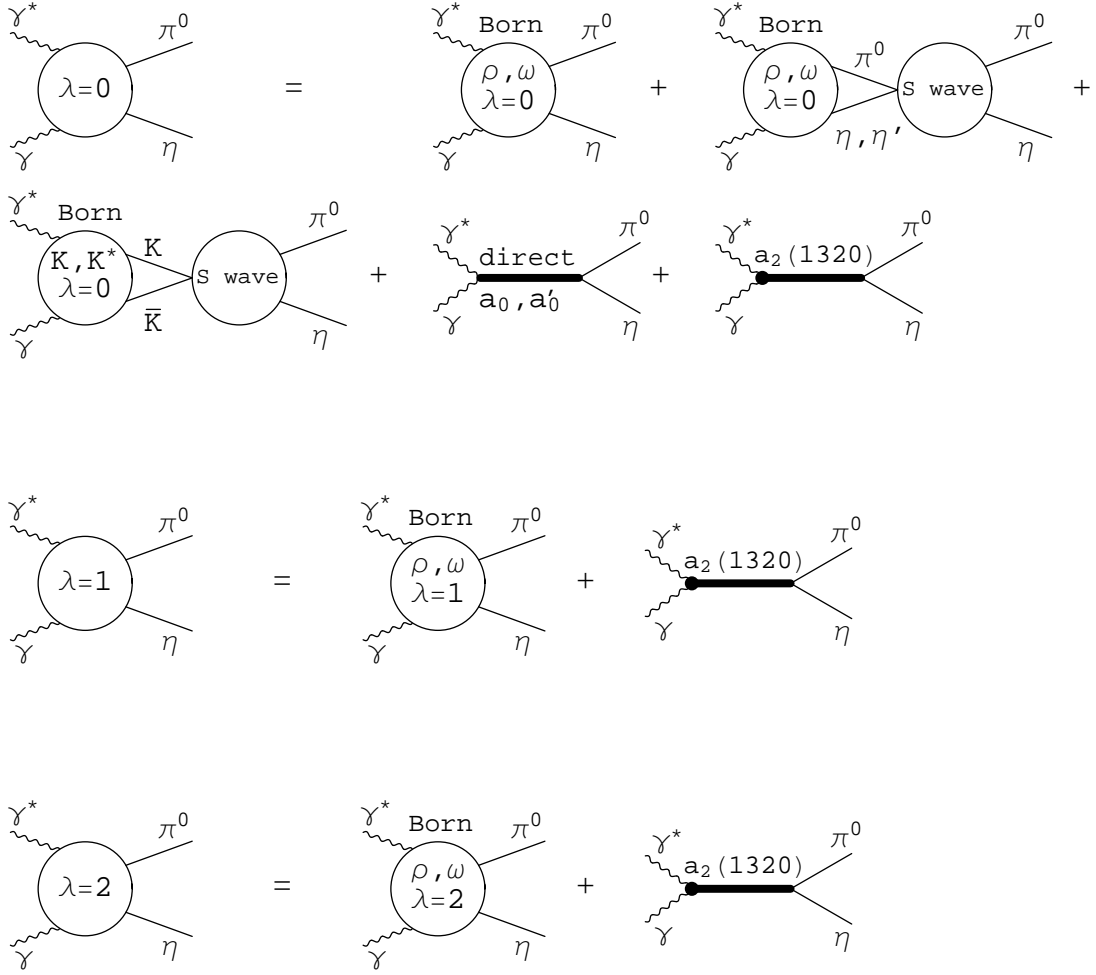}} \caption {Diagrammatical representation
for the helicity amplitudes
$\gamma^*\gamma$\,$\to$\,$\pi^0\eta$.}\label{Amplitudes}\end{figure}
\begin{figure} \begin{center}\begin{tabular}{c}{\includegraphics[width=8cm]{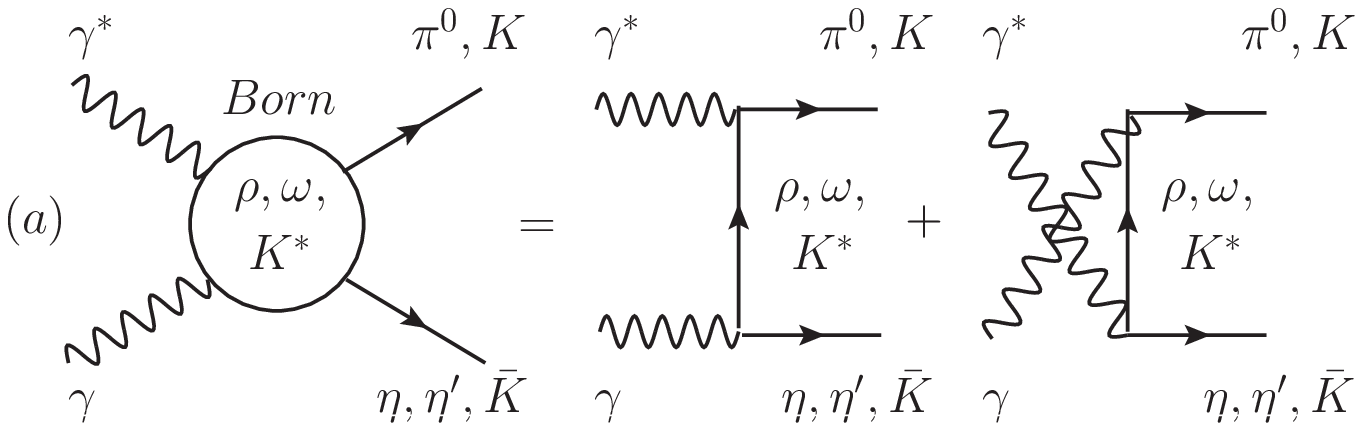}}\\
{\includegraphics[width=12cm]{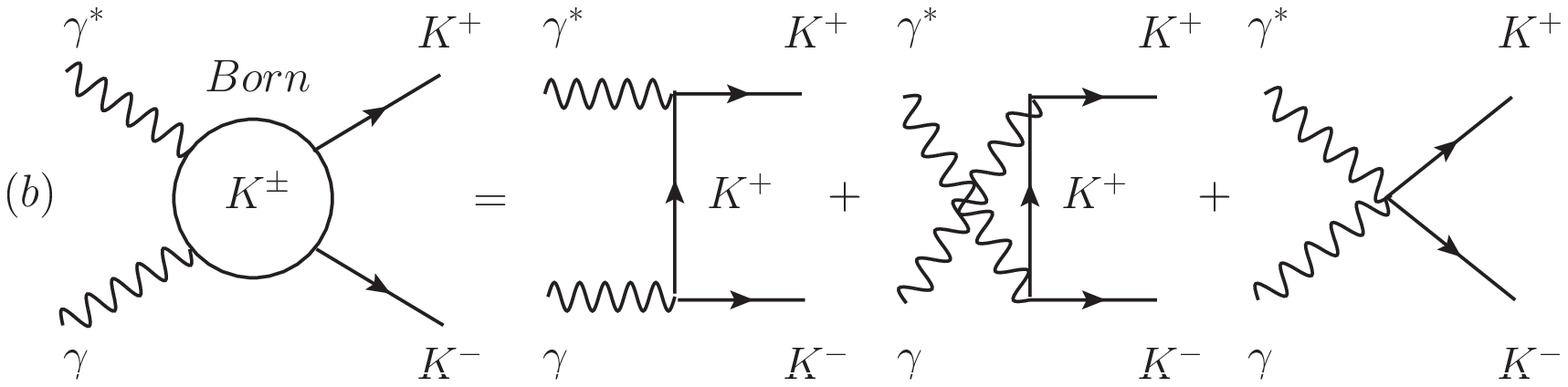}}\end{tabular}\end{center}
\vspace{-2mm} \caption {The Born $\rho$, $\omega$, $K^*$, $K$
exchange diagrams for $\gamma^*\gamma$\,$\to$\,$\pi^0\eta$,
$\gamma^*\gamma$\,$ \to$\,$\pi^0\eta'$, and
$\gamma^*\gamma$\,$\to$\,$K\bar K$. }\label{BornVAKK}\end{figure}

All formulas for the $\gamma\gamma\to \eta\pi^0$ reaction were
derived in Refs. \cite{annsgn2010,annsgn2011}, these results are
used to fit the experimental data. In this section we derive
formulas for the $\gamma^*(q)\gamma(k)\to \eta(q_1)\pi^0(q_2)$
reaction, $Q^2=-q^2 > 0$. The results of Refs.
\cite{annsgn2010,annsgn2011} are reached in the limit $Q^2\to 0$.

Most of formulas in this section are modifications of Refs.
\cite{annsgn2010,annsgn2011} results. In all diagrams the vector
resonances undergo excitation in the $\gamma^*$ line resulting in
distinctive factors in the amplitudes.

According to Refs. \cite{annsgn2010,annsgn2011}, we use a model
for the helicity amplitudes $M_\lambda$ ($\lambda$ is the
difference between photon helicities), taking into account
electromagnetic Born contributions from $\rho$, $\omega$, $K^*$,
$K$ exchanges and strong elastic and inelastic final-state
interactions in $\pi^0\eta$, $\pi^0\eta'$, $K^+K^-$, and $K^0\bar
K^0$ channels, as well as the contributions due to the direct
interaction of the resonances with photons:

\begin{eqnarray}
& M_0(\gamma^*\gamma\to\pi^0\eta;s,Q,\theta)=
M^{\mbox{\scriptsize{Born}}
\,V}_0(\gamma^*\gamma\to\pi^0\eta;s,Q,\theta)+ \widetilde{I}^V_
{\pi^0\eta}(s,Q)\,T_{\pi^0\eta\to\pi^0\eta}(s) & \nonumber\\ & +
 \widetilde{I}^V_
{\pi^0\eta'}(s,Q)\,T_{\pi^0\eta'\to\pi^0\eta}(s)+\bigg(\widetilde{I}^{K^{*+}}_{K^+K^-}(s,Q)-\widetilde{I}^{K^{*0}}_{K^0\bar
K^0}(s,Q) & \nonumber\\ & +\widetilde{I}^{K^+}_{K^+K^-}(s,Q)\bigg)
T_{K^+K^-\to \pi^0\eta}(s) & \nonumber\\ & +M^{\mbox
{\scriptsize{direct}}}_{\mbox{\scriptsize{res}}}(s,Q)+
M_0(\gamma^*\gamma\to a_2(1320)\to\pi^0\eta;s,Q,\theta), &
\label{M0}
\\[0.14cm] & M_1(\gamma^*\gamma\to\pi^0\eta;s,Q,\theta)=M^{\mbox{
\scriptsize{Born}}\,V}_1(\gamma^*\gamma\to\pi^0\eta;s,Q,\theta)&
\nonumber\\ & + M_1(\gamma^*\gamma\to
a_2(1320)\to\pi^0\eta;s,Q,\theta), & \label{M1}
\\[0.14cm] & M_2(\gamma^*\gamma\to\pi^0\eta;s,Q,\theta)=M^{\mbox{
\scriptsize{Born}}\,V}_2(\gamma^*\gamma\to\pi^0\eta;s,Q,\theta)&
\nonumber\\ & + M_2(\gamma^*\gamma\to
a_2(1320)\to\pi^0\eta;s,Q,\theta), & \label{M2}
\end{eqnarray}
the diagrams corresponding to these amplitudes are shown in Figs.
\ref{Amplitudes}, \ref{BornVAKK}. Here $\theta$ is the angle
between $\pi^0$ and $\gamma$ (or $\eta$ and $\gamma^*$) momenta in
the $\gamma^*\gamma$ center-of-mass system.

The kaon loop contribution $\gamma^*\gamma\to K^+K^-\to
a_0\to\eta\pi^0$ (the term
$\widetilde{I}^{K^+}_{K^+K^-}(s,Q)T_{K^+K^-\to \pi^0\eta}(s)$) and
the $a_2$ contribution are the largest ones, but other terms are
also essential. $I$-functions are loop integrals, $T$-functions
represent rescattering amplitudes (see below).

In the case of real photons $M_1(\gamma^*\gamma\to\pi^0\eta)$
vanishes because of the electromagnetic current conservation. The
$M_0(\gamma^*\gamma\to a_2(1320)\to\pi^0\eta)$ is small at $Q=0$
according to experiment.

The cross sections $\sigma_\lambda\equiv
\sigma_\lambda(\gamma^*\gamma\to\eta\pi^0,s,Q)$ are related to the
amplitudes $M_\lambda$ as

\begin{equation}
\sigma_\lambda= \frac{\rho_{\pi\eta}(s)}{64\pi
s\rho_{\gamma^*\gamma}}\int^{0.8}_{-0.8}|M_\lambda|^2d\cos\theta\,,\hspace{5mm}
\rho_{\gamma^*\gamma}=1+Q^2/s\,,
\end{equation}

\begin{equation}
\bar \sigma_\lambda=
\frac{1}{2a}\int^{\sqrt{s}+a}_{\sqrt{s}-a}\sigma_\lambda (s')
d\sqrt{s'}\,,
\end{equation}

\noindent and the total cross section
$\sigma(\gamma^*(Q^2)\gamma\to\eta\pi^0,s)=\sigma_0+\sigma_1+\sigma_2$.
The limits $|\cos\theta|\leq 0.8$ and bin size $2a=20$ MeV were
used in the experiment Ref. \cite{uehara}, where the data on the
sum $\bar \sigma_0+\bar \sigma_2$ at $Q=0$ were presented for
different values of $\sqrt{s}$ ($\sigma_1=\bar \sigma_1=0$ at
$Q=0$).

Let us derive all the terms in Eqs. (\ref{M0}), (\ref{M1}), and
(\ref{M2}).

The Born term $M^{\mbox{\scriptsize{Born}} \,V}_\lambda$ is caused
by equal \cite{equalBorn} contributions of the $\rho$ and $\omega$
exchange mechanisms \cite{AS88}. It is calculated by Vector
Dominance Model (VDM), the result is the significant modification
of Eqs. (3) and (4) of Ref. \cite{annsgn2010}:
\begin{eqnarray}\label{MBornV0} && \mbox{\qquad\ \ \ }
M^{\mbox{\scriptsize{Born}}
\,V}_0(\gamma^*\gamma\to\pi^0\eta;s,Q,\theta)
=2g_{\omega\pi\gamma}g_{\omega\eta\gamma}F^{\mbox{\scriptsize{Born}}}_{\eta\pi^0}(Q)\bigg[
\frac{A_0(s,t,Q,m_\eta,m_\pi)G_\omega(s,t)}{t-m^2_\omega}
\nonumber
\\[0.14cm] && +\frac{A_0(s,u,Q,m_\pi,m_\eta)G_\omega(s,
u)}{u-m^2_\omega}\bigg],\mbox{\ } \\[0.14cm] \label{MBornV1} &&
\mbox{\qquad\ \ \ } M^{\mbox{\scriptsize{Born}}
\,V}_1(\gamma^*\gamma\to\pi^0\eta;s,Q,\theta)
=2g_{\omega\pi\gamma}g_{\omega\eta\gamma}F^{\mbox{\scriptsize{Born}}}_{\eta\pi^0}(Q)\bigg[
\frac{A_1(s,t,Q,m_\eta,m_\pi)G_\omega(s,t)}{t-m^2_\omega}
\nonumber
\\[0.14cm] && +\frac{A_1(s,u,Q,m_\pi,m_\eta)G_\omega(s,
u)}{u-m^2_\omega}\bigg],\mbox{\ } \\[0.14cm] && \mbox{\qquad\ \ \
}M^{\mbox{\scriptsize{Born}}
\,V}_2(\gamma^*\gamma\to\pi^0\eta;s,Q,\theta)
=\frac{2g_{\omega\pi\gamma}g_{\omega\eta\gamma}}{4}F^{\mbox{\scriptsize{Born}}}_{\eta\pi^0}(Q)\bigg(m_\eta^2
m_\pi^2 - tu
        \nonumber\\ \label{MBornV2} && +\frac{Q^2}{s + Q^2}(t -
              m_\pi^2)(u - m_\eta^2)\bigg)\left[
\frac{G_\omega(s,t)}{t-m^2_\omega}+\frac{G_\omega(s,
u)}{u-m^2_\omega}\right],\mbox{\ } \end{eqnarray}

\noindent where

$$A_0(s,t,Q,m_1,m_2)=\frac{t(s + Q^2)}{4} +
\frac{Q^2(t-m_2^2)^2}{4(s + Q^2)}\,, $$

$$A_1(s,t,Q,m_1,m_2)=Q\sqrt{-t + \frac{(m_2^2 - m_1^2 -
Q^2)^2}{4s}}\,,$$

\noindent $t$ and $u$ are the Mandelstam variables for the
reaction $\gamma^*\gamma\to\eta\pi^0$:

$$t =
  m_\pi^2 - \frac{1 +
          Q^2/s}{2}\bigg(s + m_\pi^2 - m_\eta^2 -
            s\,\rho_{\eta\pi^0}\cos\theta\bigg)\,,$$

$$u =
  m_\eta^2 - \frac{1 +
          Q^2/s}{2}\bigg(s + m_\eta^2 - m_\pi^2 +
            s\,\rho_{\eta\pi^0} \cos\theta\bigg)\,,$$

\noindent hereafter
$\rho_{ab}(s)=2p_{ab}(s)/\sqrt{s}=\sqrt{(1-m_+^2/s)(1-m_-^2/s)}$,
$m_\pm=m_a\pm m_b$ ($ab$\,=\,$\eta\pi^0$, $K^+K^-$, $K^0\bar K^0
$, $\eta'\pi^0$), where $p_{ab}$ is the modulus of the momentum of
$a$ (or $b$) particles in the s.c.m.,
$g^2_{\omega\pi\gamma}$=$12\pi\Gamma_{\omega\to\pi\gamma}
[(m^2_\omega-m^2_\pi)/(2m_\omega)]^{-3}\approx0.519$ GeV$^{-2}$,
$g^2_{\omega\eta\gamma}$=$12\pi\Gamma_{\omega\to\eta\gamma}
[(m^2_\omega-m^2_\eta)/(2m_\omega)]^{-3}\approx 1.86\times
10^{-2}$ GeV$^{-2}$ \cite{annsgn2011}. The factor
$F^{\mbox{\scriptsize{Born}}}_{\eta\pi^0}(Q)$ is due to vector
resonances in the $\gamma^*$ line and reads

\begin{equation}
F^{\mbox{\scriptsize{Born}}}_{\eta\pi^0}(Q)=
\frac{1}{2}\bigg(\frac{1}{1+Q^2/m_\rho^2}+\frac{1}{1+Q^2/m_\omega^2}\bigg).
\label{FetaPi}
\end{equation}

As in Refs. \cite{annsgn2010,annsgn2011}, we take the formfactor
$G_\omega (s,t)=G_\rho (s,t)$ of forms

\begin{equation}
G_\omega(s,t)=\exp[(t-m^2_\omega)b_\omega (s)]\,,
\label{Gfactor}
\end{equation}

\begin{equation}
b_\omega(s)=b^0_\omega+(\alpha'_\omega/4) \ln[1+(s/s_0)^4]\,,
\end{equation}

\noindent where $b^0_\omega=0,\ \alpha'_\omega=0.8$ GeV$^{-2}$,
and $s_0=1$ GeV$^2$. Form factors for the $K^*$ exchange result
from the above by substitution of $m_{K^*}$ for $m_\omega\,$,
other parameters are the same.

The function $\widetilde{I}^V_ {\pi^0\eta}(s,Q)$ reads
\begin{equation}
\widetilde{I}^V_
{\pi^0\eta}(s,Q)=\frac{s}{\pi}\int_{(m_\eta+m_{\pi})^2}^{\infty}
ds'\, \rho_{\eta\pi}(s')\frac{M^{\mbox{\scriptsize{Born}}
\,V}_{00}(\gamma^*\gamma\to\pi^0\eta;s,Q)}{s'(s'-s-i\varepsilon)},
\end{equation}

\begin{equation}
M^{\mbox{\scriptsize{Born}}
\,V}_{00}(\gamma^*\gamma\to\pi^0\eta;s,Q)=\frac{1}{2}\int_{-1}^{1}
M^{\mbox{\scriptsize{Born}}
\,V}_0(\gamma^*\gamma\to\pi^0\eta;s,Q,\theta)\,\mbox{dcos}\,\theta
.
\end{equation}

The loop functions $\widetilde{I}^V_ {\pi^0\eta'}(s,Q)$,
$\widetilde{I}^{K^{*+}}_{K^+K^-}(s,Q)$ and
$\widetilde{I}^{K^{*0}}_{K^0\bar K^0}(s,Q)$ are built in a similar
way. In the Born amplitudes for the $\gamma^*\gamma\to\pi^0\eta'$
process the constant
$g^2_{\omega\eta'\gamma}$=$4\pi\Gamma_{\eta'\to\omega\gamma}
[(m^2_{\eta'}-m^2_\omega)/(2m_\omega)]^{-3}\approx 1.86\times
10^{-2}$ GeV$^{-2}$ \cite{pdg-2014}, also we have
$F^{\mbox{\scriptsize{Born}}}_{\eta'\pi^0}(Q)=F^{\mbox{\scriptsize{Born}}}_{\eta\pi^0}(Q)$.
The amplitudes
$M^{\mbox{\scriptsize{Born}}\,K^*}_\lambda(\gamma^*\gamma \to
K\bar K;s,\theta)$ for the $K^*$ exchanges result from Eqs.
(\ref{MBornV0}), (\ref{MBornV1}), (\ref{MBornV2}) with the help of
substitutions $m_\omega$\,$\to$\,$m_{K^*}$,
$G_\omega$\,$\to$\,$G_{K^*}$, $m_\pi$\,$\to$\,$m_K$,
$m_\eta$\,$\to$\,$m_K$, and
$2g_{\omega\pi\gamma}g_{\omega\eta\gamma}$\,$\to$\,$g^2_{K^*K\gamma}$,
where $g^2_{K^{*+}K^+\gamma}\approx0.064$ GeV$^{-2}$ and
$g^2_{K^{*0} K^0\gamma} \approx 0.151$ GeV$^{-2}$
\cite{annsgn2011}. The factors $F^{\mbox{\scriptsize{Born}}\, K^{*
+}}_{KK}(Q)$ and $F^{\mbox{\scriptsize{Born}}\, K^{* 0}}_{KK}(Q)$,
provided by quark counting, are

\begin{equation}
F^{\mbox{\scriptsize{Born}}\, K^{* +}}_{KK}(Q)=
\frac{3/2}{1+Q^2/m_\rho^2}+\frac{1/2}{1+Q^2/m_\omega^2}-\frac{1}{1+Q^2/m_\phi^2}\,,
\label{fKKstar}
\end{equation}

\begin{equation}
F^{\mbox{\scriptsize{Born}}\, K^{* 0}}_{KK}(Q)=
\frac{3/4}{1+Q^2/m_\rho^2}-\frac{1/4}{1+Q^2/m_\omega^2}+\frac{1/2}{1+Q^2/m_\phi^2}\,.
\label{fKKstar0}
\end{equation}

The kaon loop integral $\widetilde{I}^{K^+}_{K^+K^-}(s,Q)$ is

$$\widetilde{I}^{K^+}_{K^+K^-}(s,Q)=-F_{K^+}(Q)8\alpha\bigg\{1 +
\frac{Q^2}{s + Q^2}\bigg(\rho_{K^+K^-}(s)\Big(\ln\frac{1 +
\rho_{K^+K^-}(s)}{1 - \rho_{K^+K^-}(s)} - i\pi\Big) -$$

$$\rho_{K^+K^-}(-Q^2)\ln\frac{1 + \rho_{K^+K^-}(-Q^2)}{
\rho_{K^+K^-}(-Q^2)-1} + $$

\begin{equation}
\frac{m_{K^+}^2}{Q^2}\Big(-\ln^2\frac{1 +
\rho_{K^+K^-}(-Q^2)}{\rho_{K^+K^-}(-Q^2)-1} + (\ln\frac{1 +
\rho_{K^+K^-}(s)}{1 - \rho_{K^+K^-}(s)} -
                        i\pi)^2\Big)\bigg)\bigg\} ,
                        \label{IKK}
\end{equation}

\begin{equation}
F_{K^+}(Q)=
\frac{1/2}{1+Q^2/m_\rho^2}+\frac{1/6}{1+Q^2/m_\omega^2}+\frac{1/3}{1+Q^2/m_\phi^2}
. \label{Fkp}
\end{equation}

The amplitudes of the pseudoscalar pairs rescattering are

\begin{eqnarray} && \mbox{\qquad}
T_{\pi^0\eta\to\pi^0\eta}(s)=T_0^1 (s)= \frac{\eta^1_0(s)e^{2i
\delta^1_0(s)}-1}{2i\rho_{\pi\eta}(s)}=T_{\pi\eta}^{bg}(s)+e^{2i\delta_{\pi\eta}^{bg}
(s)}T^{res}_{\pi^0\eta\to\pi^0\eta}(s)\,, \label{Tpietapieta}\\ &&
T_{\pi^0\eta'\to\pi^0\eta}(s)=T^{res}_{\pi^0\eta'
\to\pi^0\eta}(s)\,e^{i[\delta_{\pi\eta'}^{bg}(s)+\delta_{\pi\eta}^{bg}
(s)]},\label{Tpieta1pieta} \\ && T_{K^+K^-\to\pi^0\eta}(s)=
T^{res}_{K^+K^-\to\pi^0 \eta}(s)\,e^{i[\delta_{K\bar
K}^{bg}(s)+\delta_{\pi\eta}^{bg}(s)]},\ \ \ \ \label{TKKpieta}
\end{eqnarray} where
$T_{\pi\eta}^{bg}(s)=(e^{2i\delta_{\pi\eta}^{bg}(s)}-1)/(2i
\rho_{\pi\eta}(s))$, $T^{res}_{\pi^0\eta\to\pi^0\eta}(s)
=(\eta^1_0(s)e^{2i\delta_{\pi\eta}^{res}(s)}-1)/(2i\rho_{
\pi\eta}(s))$, $\delta^1_0(s)=\delta_{\pi\eta}^{bg}(s)+
\delta_{\pi\eta}^{res}(s)$, $\delta_{\pi\eta}^{bg}(s)$,
$\delta_{\pi\eta'}^{bg}(s)$, and $\delta_{K\bar K}^{bg}(s)$ are
the phase shifts of the elastic background contributions in the
channels $\pi\eta$, $\pi\eta'$, and $K\bar K$ with isospin $I=1$,
respectively.

When $a_0'$ is taken into account the resonant amplitudes of the
processes $ab\to \eta\pi^0$ are

\begin{equation}
T^{res}_{ab\to
\eta\pi^0}(s)=\sum_{R,R'}\frac{g_{Rab}G_{RR'}^{-1}g_{R'\eta\pi^0}}{16\pi}\,,\label{Tres}\end{equation}

\noindent where $R,R'=a_0,a_0'$ and pair
$ab=\gamma\gamma,\,\eta\pi^0,\,K^+K,\,\eta'\pi^0$.

For $a,b\neq \gamma\gamma$ the constants  $g_{Rab}$ are related to
the width
\begin{equation}
\Gamma_R(m)=\sum_{ab} \Gamma(R\to
ab,m)=\sum_{ab}\frac{g_{Rab}^2}{16\pi m}\rho_{ab}(m).
\label{f0pipi}
\end{equation}

In case of $ab=\gamma\gamma$ the constants
$g_{R\gamma\gamma}\equiv g^{(0)}_{R\gamma\gamma}$ are related to
the "direct" width as

\begin{equation}
\Gamma^{(0)}_{R\to\gamma\gamma}=\frac{|m_R^2
g^{(0)}_{R\gamma\gamma}|^2}{16\pi m_R}
\,.\label{GammaDirect}\end{equation}

Remind that this is only a part of $R\to\gamma\gamma$ width, which
is mainly produced by rescatterings.

The matrix of the inverse propagators \cite{achasov-97} is

\[G_{RR'}\equiv G_{RR'}(m)=\left( \begin{array}{cc} D_{a_0'}(m)&-\Pi_{a_0'a_0}(m)\\-\Pi_{a_0'a_0}(m)&D_{a_0}(m)\end{array}\right),\]

$$\Pi_{a'_0a_0}(m)=\sum_{a,b} \frac{g_{a_0' ab}}{g_{a_0
ab}}\Pi^{ab}_{a_0}(m)+C_{a'_0a_0},$$

\noindent where $m$ is the invariant mass of the $\eta\pi^0$
system, $m^2=s$, the constant $C_{a_0'a_0}$ incorporates the
subtraction constant for the transition $a_0(980)\to(0^-0^-)\to
a_0'$ and effectively takes into account contribution of
multi-particle intermediate states to $a_0\leftrightarrow a_0'$
transition, see Ref. \cite{achasov-97}. The inverse propagator of
the R scalar meson is presented also in Refs.
\cite{achasov-89,achasov-97}:

\begin{equation}
\label{propagator} D_R(m)=m_R^2-m^2+\sum_{ab}[Re
\Pi_R^{ab}(m_R^2)-\Pi_R^{ab}(m^2)],
\end{equation}
\noindent where $\sum_{ab}[Re \Pi_R^{ab}(m_R^2)-
\Pi_R^{ab}(m^2)]=Re\Pi_R(m_R^2)- \Pi_R(m^2)$ takes into account
the finite width corrections of the resonance which are the one
loop contribution to the self-energy of the $R$ resonance from the
two-particle intermediate  $ab$ states.

Polarization operators of the $a_0$ and $a_0'$ are provided in
Appendix I.

For the background phase shifts we use the parametrizations from
Refs. \cite{annsgn2010,annsgn2011}:
\begin{equation}
e^{2i\delta^{bg}_{ab}(s)}=\frac{1+iF_{ab}(s)}{1-iF_{ab}(s)},
\end{equation} where
\begin{eqnarray}
F_{\pi\eta}(s)=\frac{\sqrt{1-(m_\eta+m_\pi)^2/s}\left(c_0+c_1
\left(s-(m_\eta+m_\pi)^2\right)\right)}
{1+c_2\left(s-(m_\eta+m_\pi)^2\right)^2}\,, && \\ F_{K\bar
K}(s)=f_{K\bar K}\sqrt{s-4m_{K^+}^2}\,, &&
\\ F_{\pi\eta'}(s)=f_{\pi\eta'}\sqrt{s-(m_{\eta'}+m_\pi)^2}\,.
\mbox{\qquad\qquad} && \end{eqnarray}

Note that analytical continuation of the phases under the
thresholds changes modules of corresponding amplitudes. The
parameterization of $F_{K\bar K}(s)$ slightly differs from Refs.
\cite{annsgn2010,annsgn2011}.

The amplitude $M^{\mbox{\scriptsize{direct}}}_
{\mbox{\scriptsize{res}}}(s,Q)$:
\begin{equation}\label{Mdirect}
M^{\mbox{\scriptsize{direct}}}_ {\mbox{\scriptsize{res}}}(s,Q)=
16\pi s\, T^{res}_{\gamma\gamma\to
\eta\pi^0}(s)F^{\mbox{\scriptsize{direct}}}(s,Q)
e^{i\delta_{\pi\eta}^{bg}(s)} ,
\end{equation}

\begin{equation}
F^{\mbox{\scriptsize{direct}}}(s,Q)=\frac{1}{2}\bigg(\frac{1}{1+Q^2/m_\rho^2}+\frac{1}{1+Q^2/m_\omega^2}\bigg)(1+Q^2/s)
, \label{Fdirect}
\end{equation}

\noindent describes the $\gamma^*\gamma\to\pi^0\eta$ transition
caused by the direct coupling constants of the $a_0$ and $a'_0$
resonances to photons $g^{(0)}_{a_0\gamma\gamma}$ and
$g^{(0)}_{a'_0\gamma\gamma }$. $T^{res}_{\gamma\gamma\to
\eta\pi^0}$ is defined in Eqs. (\ref{Tres}) and
(\ref{GammaDirect}), the factor $s(1+Q^2/s)$ appears due to the
gauge invariance (the effective Lagrangian $\sim
F_{\mu\nu}F^{\mu\nu}a_0$, $M^{\mbox{\scriptsize{direct}}}_
{\mbox{\scriptsize{res}}}(s,Q)\sim (kq)$).

It is known that in the reaction $\gamma\gamma \to a_2\to \eta\pi$
tensor mesons are produced mainly by the photons with the opposite
helicity states. The effective Lagrangian in this case is

\begin{equation}
L=g_{a_2 \gamma\gamma}T_{\mu\nu}F_{\mu\sigma}F_{\nu\sigma}\,,\label{lagra2gg}
\end{equation}
$$F_{\mu\sigma}=\partial_\mu A_\sigma - \partial_\sigma A_\mu\,,
$$

\noindent where $A_{\mu}$ is a photon field and $T_{\mu\nu}$ is a
tensor $a_2$ field. So in the frame of Vector Dominance Model we
assume that the effective Lagrangian of the reaction $a_2\to
V(1)V(2)$ is \cite{akar-86}:

\begin{equation}
L=g_{a_2 V(1) V(2)}T_{\mu\nu}F^{V(1)}_{\mu\sigma}F^{V(2)}_{
\nu\sigma}\,,\label{lagra2vv}
\end{equation}
$$F^{V(i)}_{\mu\sigma}=\partial_\mu V(i)_{\sigma}-\partial_\sigma
V(i)_{\mu} \, ;\, i= 1, 2\,.$$

Matrix elements for $a_2(1320)$ contribution are in Ref.
\cite{akar-86}:

\begin{equation}
M_2(\gamma^*\gamma\to
a_2(1320)\to\pi^0\eta;s,Q,\theta)=A(s,Q)\sin^2\theta ,
\label{matA2lambda2}
\end{equation}

\begin{equation}
M_1(\gamma^*\gamma\to
a_2(1320)\to\pi^0\eta;s,Q,\theta)=-\sqrt{2}A(s,Q)\sqrt{\frac{Q^2}{s}}\sin\theta\cos\theta
, \label{matA2lambda1}
\end{equation}

\begin{equation}
M_0(\gamma^*\gamma\to
a_2(1320)\to\pi^0\eta;s,Q,\theta)=-A(s,Q)\frac{Q^2}{s}\Big(\cos^2\theta
-\frac{1}{3}\Big) , \label{matA2lambda0}
\end{equation}

\begin{equation}
A(s,Q)=20\pi F_{a_2}(Q)\sqrt{\frac{6
s\Gamma_{a_2\to\gamma\gamma}(s)\Gamma_{a_2\to\eta\pi^0}(s)}{\rho_{\eta\pi^0}(s)}}\,\frac{1}{D_{a_2}(s)}\,\bigg(1+\frac{Q^2}{s}\bigg)
. \label{commonFactorA2}
\end{equation}

The $F_{a_2}(Q)$ is defined and discussed in Sec.
\ref{vectorExcitations}. The $F_{a_2}(0)=1$, and the $F_{a_2}$
dependence on $Q$ does not influence on the
$\gamma\gamma\to\eta\pi^0$ process. Note that Eq.
(\ref{commonFactorA2}) differs from Eq. (3) in Ref. \cite{akar-86}
because of normalization of the amplitude and the sign of
$g_{a_2\gamma\gamma}$, which is taken negative.

As in Refs. \cite{annsgn2010,annsgn2011} we take

\begin{equation} D_{a_2}(m^2)
=m_{a_2}^2-m^2-im\Gamma_{a_2}(m)\,,\end{equation}

\noindent where

\begin{equation}\Gamma_{a_2}(m)=
\Gamma_{a_2}^{tot}\frac{m_{a_2}^2}{m^2}\frac{p^5_{\eta\pi}(m)}
{p^5_{\eta\pi}(m_{a_2})}\frac{D_2(r_{a_2}p_{\eta\pi}(m_{a_2}))}{D_2(r_{a_2}p_{\eta\pi}(m))}\,,
\end{equation} \noindent and

$$\Gamma_{a_2\to\eta\pi^0}(m)=Br(a_2\to\eta\pi^0)\Gamma_{a_2}(m) .
$$

\noindent

Here $D_2(x)=9+3x^2+x^4$ \cite{Blatt-Weisskopf}, and we take
$m_{a_2}=1318.3$ MeV, $\Gamma_{a_2}^{tot}=105$ MeV, and
$Br(a_2\to\eta\pi^0)=0.145$ from Ref. \cite{pdg-2014}.
$\Gamma_{a_2\to\gamma\gamma}(s)=(\sqrt{s}/m_{a_2})^3\Gamma_{a_2\to\gamma\gamma}(m_{a_2})$,
$Br(a_2\to\gamma\gamma)=9.4\times 10^{-6}$ \cite{pdg-2014}. We
also take $r_{a_2}=1.9$ GeV$^{-1}$ from Refs.
\cite{annsgn2010,annsgn2011}.

The $\phi\to\eta\pi^0\gamma$ decay description. Theoretical
description of the KLOE data Ref. \cite{kloea0} on the mass
spectrum $dBr(\phi\to\gamma\pi^0\eta,m)/dm$ is the same as in Ref.
\cite{our_a0}, with obvious change

\begin{equation}
\frac{g_{a_0K^+K^-}g_{a_0\eta\pi^0}}{D_{a_0}(m)} \to
\sum_{R,R'}g_{RK^+K^-}G_{RR'}^{-1}g_{R'\eta\pi^0}\,.
\end{equation}

The resulted formulas are in Appendix II.

\section{Results of data description}
\label{results}

\begin{figure}
\begin{center}
\begin{tabular}{ccc}
\includegraphics[width=8cm]{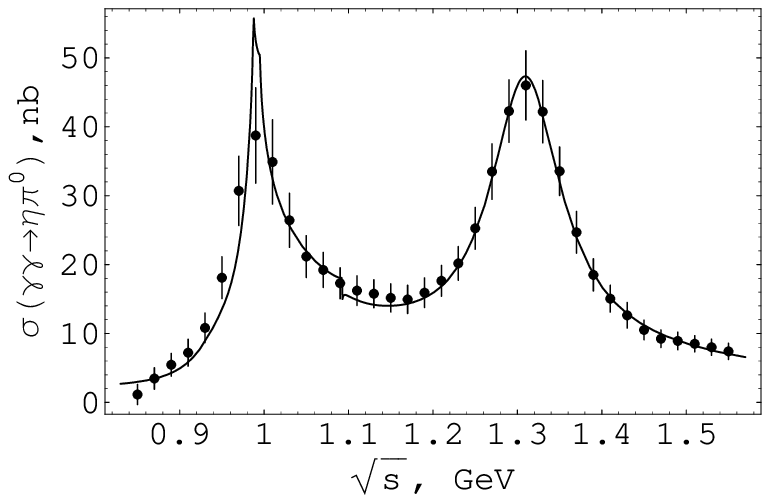}& \includegraphics[width=8cm]{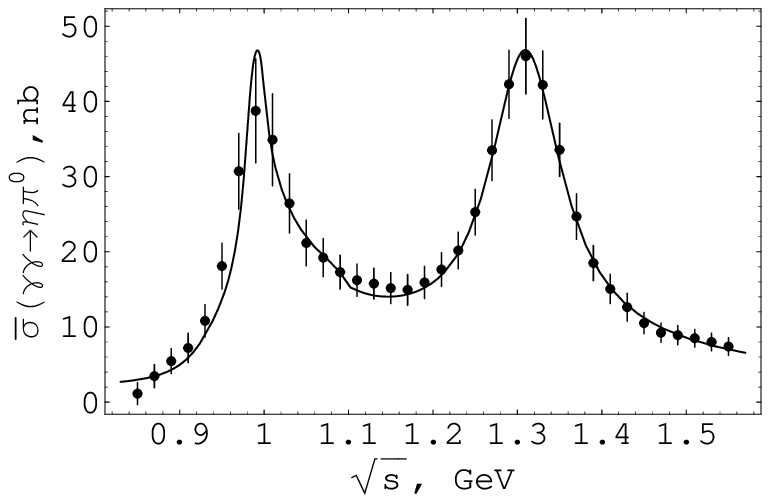}\\ (a)&(b)
\end{tabular}
\end{center}
\caption{The $\gamma\gamma\to \eta\pi^0$ cross section,
$|\cos\theta|<0.8$. The curves correspond to Fit 1, points are the
Belle data \cite{uehara}. The curve on the (a) figure represents
cross section as is, while the curve on the (b) figure represents
averaged cross section: each point of the curve is the cross
section averaged over the $\pm 10$ MeV neighborhood. }
\label{gammaGamma1400}
\end{figure}

\begin{figure}
\begin{center}
\includegraphics[width=8cm]{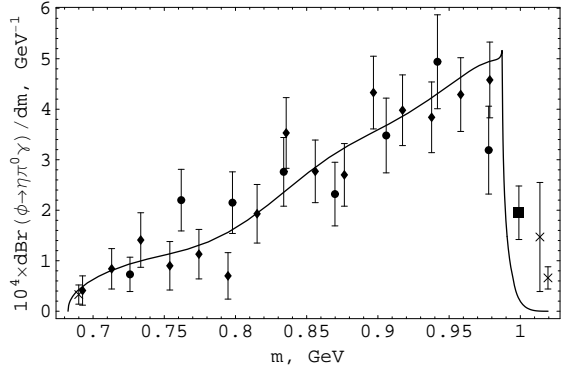}
\end{center}
\caption{Plot of the Fit 1 curve and the KLOE data (points)
\cite{kloea0} on the $\phi\to\eta\pi^0\gamma$ decay. Cross points
are omitted in fitting.}\label{phiSpectrum}
\end{figure}

\begin{figure}
\begin{center}
\begin{tabular}{ccc}
\includegraphics[width=8cm]{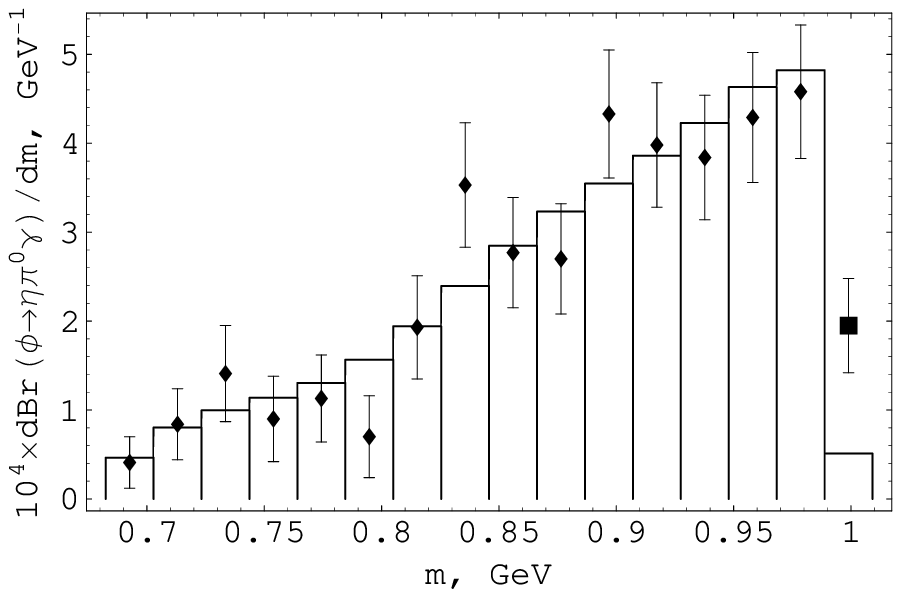}& \includegraphics[width=8cm]{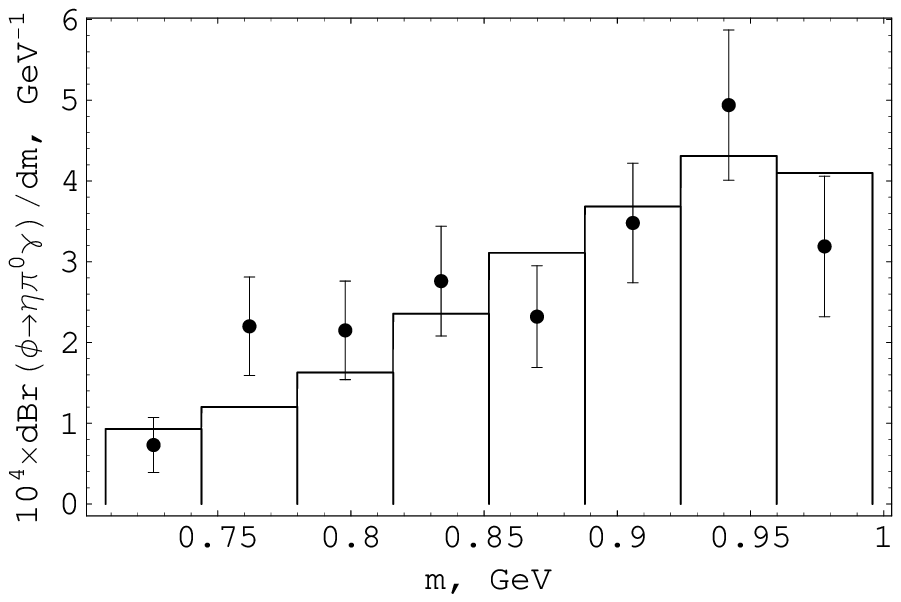}\\ (a)&(b)
\end{tabular}
\end{center}
\caption{The comparison of the KLOE data on
$\phi\to\eta\pi^0\gamma$ decay and Fit 1. Histograms show Fit 1
curve averaged over each bin for (a) $\phi\to\eta\pi^0\gamma$,
$\eta\to\gamma\gamma$ and (b) $\phi\to\eta\pi^0\gamma$,
$\eta\to\pi^+\pi^-\pi^0$ samples, see details in Ref.
\cite{our_a0}.} \label{phiSpectrumBins}
\end{figure}

Using the above theoretical framework we fit the data Refs.
\cite{uehara,kloea0} and obtain results shown in Table I and Figs.
\ref{gammaGamma1400}, \ref{phiSpectrum}, and
\ref{phiSpectrumBins}.

In Fit 1 the $m_{a'_0}=1400$ MeV is in the middle of agreed
corridor $\approx 1300-1500$ MeV \cite{pdg-2014}, the $a_0$
coupling constant relations are close to the naive four-quark
model predictions \cite{achasov-89}:

\begin{eqnarray}
& g_{a_0\eta\pi^0}=\sqrt{2}
\mbox{sin}(\theta_p+\theta_q)g_{a_0K^+K^-}=(0.85\div 0.98)
g_{a_0K^+K^-} , & \nonumber\\ & g_{a_0\eta'\pi^0}=-\sqrt{2}
\mbox{cos}(\theta_p+\theta_q)g_{a_0K^+K^-}=-(1.13\div 1.02)
g_{a_0K^+K^-} . & \label{FourQrelations}
\end{eqnarray}

In brackets first values correspond to $\theta_p=-18^\circ$ and
the second ones to $\theta_p=-11^\circ$. The
$\theta_q=54.74^\circ$.

In Fit 1 the $a_0$ and $a_0'$ direct couplings with the
$\gamma\gamma$ channel are close to their values in Refs.
\cite{annsgn2010,annsgn2011}.

One can see that the quality of experimental data description is
good, see also Figs. \ref{gammaGamma1400}, \ref{phiSpectrum}, and
\ref{phiSpectrumBins}. This means that the data agree with the
four-quark model scenario.

Fits 1, 3, and 4 are obtained with some restrictions, see Appendix
III for details. Fit 2 is obtained without these restrictions by
minimization of the $\chi^2$ function only at $m_{a_0'}=1400$ MeV.
It shows that in the absence of restrictions the $a_0$ coupling
constants go not too far from the four-quark model prediction and
kaon loop gives main contribution to the $a_0\to\gamma\gamma$
width.

In Fits 3 and 4 the $a_0'$ mass $m_{a_0'}$ is set to $1300$ MeV
and $1500$ MeV, respectively. Fits 3 and 4 show that the
experimental data under consideration allow large range of
$m_{a_0'}$ values. Note that it is possible to obtain good fits
with $m_{a_0'}=1200$ MeV and $m_{a_0'}=1700$ MeV also
\cite{a0prime}.

Note also that both the data on $\gamma\gamma\to \eta\pi^0$ and
the data on $\phi\to\eta\pi^0\gamma$ decay can be described
without $a_0'$ contribution. But good $\gamma\gamma\to \eta\pi^0$
description (we achieved $\chi^2/n.d.f.=15.1/27$) requires large
width of the $a_0(980)$ ($\Gamma_{a_0}(m_{a_0})\approx 650$ MeV,
$\Gamma^{eff}_{a_0}\approx 100$ MeV), which contradicts the data
on the $\phi\to\eta\pi^0\gamma$ decay (see also Ref.
\cite{our_a0}).

The $g^{(0)}_{a_0\gamma\gamma}$ is caused by the $q\bar q$
component of the $a_0(980)$, $a^0_{2q}=(u\bar u - d\bar
d)/\sqrt{2}$, while in the four-quark model \cite{jaffe}
$a^0_{4q}=(us\bar u \bar s - ds\bar d \bar s)/\sqrt{2}$ decays to
$\gamma\gamma$ only via loops. The obtained values of
$g^{(0)}_{a_0\gamma\gamma}$ mean that the $\phi a_0\gamma$
point-like coupling constant $g_{\phi a_0\gamma}$ is negligible:
the $g_{a_0\omega\gamma}\approx
g^{(0)}_{a_0\gamma\gamma}\frac{f_\omega}{e}$, and
$g_{a_0\phi\gamma}$ should be less than $g_{a_0\omega\gamma}$ at
$\sim 20$ times because of $\phi-\omega$ mixing. Special fit with
point-like contributions $g_{\phi a_0\gamma}$ and $g_{\phi
a_0'\gamma}$ confirmed that it is not possible to extract these
constants from the current data.

Note that the inverse sign of $g_{a_2\gamma\gamma}$ does not lead
to essential consequences.

\newpage

\begin{center}
Table I. Properties of the resonances and main characteristics
\begin{tabular}{|c|c|c|c|c|c|}\hline

Fit & 1 & 2 & 3 & 4 \\ \hline

$m_{a_0}$, MeV & $993.9$ & $994.5$ & $995.3$ & $988.7$ \\ \hline

$g_{a_0K^+K^-}$, GeV  & $2.75$ & $2.43$ & $3.87$ & $3.71$
\\ \hline

$g_{a_0K^+K^-}^2/4\pi$, GeV$^2$ & $0.60$ & $0.47$ & $1.19$ &
$1.10$
\\ \hline

$g_{a_0 \eta\pi}$, GeV  & $2.74$ & $3.09$ & $3.69$ & $3.61$
\\ \hline

$g_{a_0 \eta\pi}^2/4\pi$, GeV$^2$ & $0.60$ & $0.76$ & $1.09$ &
$1.04$
\\ \hline

$g_{a_0 \eta'\pi}$, GeV  & $-2.86$ & $-4.62$ & $-4.15$ & $-3.95$
\\ \hline

$g_{a_0 \eta'\pi}^2/4\pi$, GeV$^2$ & $0.65$ & $1.70$ & $1.37$ &
$1.24$
\\ \hline

$g^{(0)}_{a_0 \gamma\gamma}\,$, $10^{-3}\,$GeV$^{-1}$ & $1.8$ &
$2.7810$ & $4.5919$ & $3.2520$
\\ \hline

$\Gamma^{(0)}_{a_0\to\gamma\gamma}$, keV & $0.063$ & $0.151$ &
$0.414$ & $0.203$
\\ \hline

$< \Gamma^{direct}_{a_0\to\gamma\gamma}>_{\eta\pi^0}$, keV &
$0.019$ & $0.031$ & $0.030$ & $0.024$
\\ \hline

$< \Gamma_{a_0\to (K\bar K + \eta\pi^0 +
\eta'\pi^0)\to\gamma\gamma}>_{\eta\pi^0}$, keV & $0.13$ & $0.11$ &
$0.14$ & $0.13$
\\ \hline

$< \Gamma_{a_0\to (K\bar K + \eta\pi^0 +
\eta'\pi^0+direct)\to\gamma\gamma}>_{\eta\pi^0}$, keV & $0.23$ &
$0.24$ & $0.27$ & $0.24$
\\ \hline

$\Gamma_{a_0}(m_{a_0})$, MeV & $116.8$ & $140.8$ & $218.7$ &
$186.8$
\\ \hline

$\Gamma^{eff}_{a_0}$, MeV & $34.6$ & $57.2$ & $38.8$ & $44.8$
\\ \hline

$m_{a'_0}$, MeV & $1400$ & $1400$ & $1300$ & $1500$ \\ \hline

$g_{a_0' K^+K^-}$, GeV  & $1.63$ & $0.26$ & $0.93$ & $2.36$
\\ \hline

$g_{a_0' K^+K^-}^2/4\pi$, GeV$^2$ & $0.21$ & $0.005$ & $0.07$ &
$0.44$ \\ \hline

$g_{a_0' \eta\pi}$, GeV  & $-3.12$ & $-1.49$ & $-2.55$ & $-2.82$
\\ \hline

$g_{a_0' \eta\pi}^2/4\pi$, GeV$^2$ & $0.77$ & $0.18$ & $0.52$ &
$0.63$ \\ \hline

$g_{a_0' \eta'\pi}$, GeV  & $-4.75$ & $-7.19$ & $-5.20$ & $-5.77$
\\ \hline

$g_{a_0' \eta'\pi}^2/4\pi$, GeV$^2$ & $1.80$ & $4.12$ & $2.15$ &
$2.64$
\\ \hline

$g_{a'_0 \gamma\gamma}$, $10^{-3}\,$GeV$^{-1}$ & $5.5$ & $9.927$ &
$8.2322$ & $8.8103$
\\ \hline

$\Gamma^{(0)}_{a'_0\to\gamma\gamma}(m_{a'_0})$, keV & $1.7$ &
$5.4$ & $3.0$ & $5.2$
\\ \hline

$\Gamma_{a'_0}(m_{a'_0})$, MeV & $330.9$ & $399.4$ & $271.0$ &
$453.4$
\\ \hline

$C_{a_0 a'_0}$, GeV$^2$ & $0.021$ & $0.302$ & $-0.021$ & $-0.034$
\\ \hline

\end{tabular}
\end{center}

\newpage
\begin{center}
Table I (continuation).

\begin{tabular}{|c|c|c|c|c|c|}\hline

Fit & 1 & 2 & 3 & 4 \\ \hline

$c_0$ & $10.3$ & $238.0$ & $31.6$ & $9.3$ \\ \hline

$c_1$, GeV$^{-2}$ & $-24.2$ & $-554.8$ & $-76.6$ & $-22.9$
\\ \hline

$c_2$, GeV$^{-4}$ & $-0.0009$ & $96.1$ & $14.6$ & $-0.0011$
\\ \hline

$f_{K\bar K}$, GeV$^{-1}$ & $-0.506$ & $-0.155$ & $0.671$ &
$0.456$ \\ \hline

$f_{\pi\eta'}$, GeV$^{-1}$ & $27.0$ & $100.2$ & $1.9$ & $50.4$
\\ \hline

$\delta,^{\circ}$ & $-94.5$ & $-132.0$ & $-166.3$ & $-144.3$
\\ \hline

$\chi^2_{\gamma\gamma}$ / $36$ points & $12.4$ & $4.8$ & $5.3$ &
$6.7$
\\ \hline

$\chi^2_{sp}$ / $24$ points & $24.5$ & $24.7$ & $24.1$ & $24.3$
\\ \hline

($\chi^2_{\gamma\gamma}$+$\chi^2_{sp}$)/n.d.f. & $36.9/46$ &
$29.5/46$ & $29.4/46$ & $31.0/46$
\\ \hline

\end{tabular}
\end{center}

Remind that there should be no confusion due to relatively large
$a_0$ width in Table I. The invariant mass spectrum of $\eta\pi^0$
in $a_0\to\eta\pi^0$ is given by the relation

\begin{equation}
\frac{dN_{\eta\pi^0}}{dm}\sim \frac{2m^2}{\pi}
\frac{\Gamma(a_0\to\eta\pi^0,m)}{|D_{a_0}(m)|^2}.
\end{equation}

The width of this distribution $\Gamma^{eff}_{a_0}$ is much less
then the nominal width $\Gamma (a_0\to\eta\pi^0, m_{a_0})$ due to
strong $a_0K\bar K$ coupling, usually $\Gamma^{eff}_{a_0}$ is
$50-70$ MeV, see Table I and Ref. \cite{our_a0}.

Since the $a_0\to\gamma\gamma$ amplitude changes rapidly near the
$K\bar K$ threshold, it is reasonable to determine the effective
width of the $a_0(980)\to\gamma\gamma$ decay averaged over the
resonance mass distribution in the $\eta\pi^0$ channel
\cite{AS88,annsgn2010}:

\begin{equation}\langle\Gamma_{a_0\to\gamma\gamma}\rangle_{\eta\pi^0}=
\int\limits_{0.9\mbox{\,\scriptsize{GeV}}}^{1.1\mbox{\,\scriptsize{GeV}}}
\frac{s}{4\pi^2}\sigma_0^{\mbox{\scriptsize{res}}}(\gamma\gamma\to
\pi^0\eta;s)d\sqrt{s} , \end{equation}

\noindent where
$\langle\Gamma_{a_0\to\gamma\gamma}\rangle_{\eta\pi^0}\equiv <
\Gamma_{a_0\to (K\bar K + \eta\pi^0 +
\eta'\pi^0+direct)\to\gamma\gamma}>_{\eta\pi^0}$, the integral is
taken over the region occupied by the $a_0(980)$ resonance, and
the $\sigma_0^{\mbox{\scriptsize{res}}}$ is determined by the
matrix element that contains only the resonance contributions from
the rescatterings and direct transitions in Eq. (\ref{M0}), i.e.
all contributions mentioned in Eq. (\ref{M0}) at $Q=0$ except the
Born one:

$$M_0^{\mbox{\scriptsize{res}}}(s)=\widetilde{I}^V_
{\pi^0\eta}(s,0)\,T_{\pi^0\eta\to\pi^0\eta}(s) +
 \widetilde{I}^V_{\pi^0\eta'}(s,0)\,T_{\pi^0\eta'\to\pi^0\eta}(s)+ $$

\begin{equation}
\bigg(\widetilde{I}^{K^{*+}}_{K^+K^-}(s,0)-\widetilde{I}^{K^{*0}}_{K^0\bar
K^0}(s,0) +\widetilde{I}^{K^+}_{K^+K^-}(s,0)\bigg) T_{K^+K^-\to
\eta\pi^0}(s) +M^{\mbox
{\scriptsize{direct}}}_{\mbox{\scriptsize{res}}}(s,0) .
\end{equation}

This quantity is an adequate characteristic of the coupling of the
$a_0(980)$ resonance with a $\gamma\gamma$ pair. One can also
consider particular contributions to
$\langle\Gamma_{a_0\to\gamma\gamma}\rangle_{\eta\pi^0}$. The
obtained results for $<
\Gamma^{direct}_{a_0\to\gamma\gamma}>_{\eta\pi^0}$ , $<
\Gamma_{a_0\to (K\bar K + \eta\pi^0
+\eta'\pi^0)\to\gamma\gamma}>_{\eta\pi^0}$, and $< \Gamma_{a_0\to
(K\bar K + \eta\pi^0
+\eta'\pi^0+direct)\to\gamma\gamma}>_{\eta\pi^0}$ are shown in
Table I. One can see that the averaged values are much less than
the values at $m=m_{a_0}$, for example, $<
\Gamma^{direct}_{a_0\to\gamma\gamma}>_{\eta\pi^0}$ is at $3$-$10$
times less than $\Gamma^{(0)}_{a_0\to\gamma\gamma}$ depending on
fit.

\begin{figure}
\begin{center}
\begin{tabular}{ccc}
\includegraphics[width=8cm]{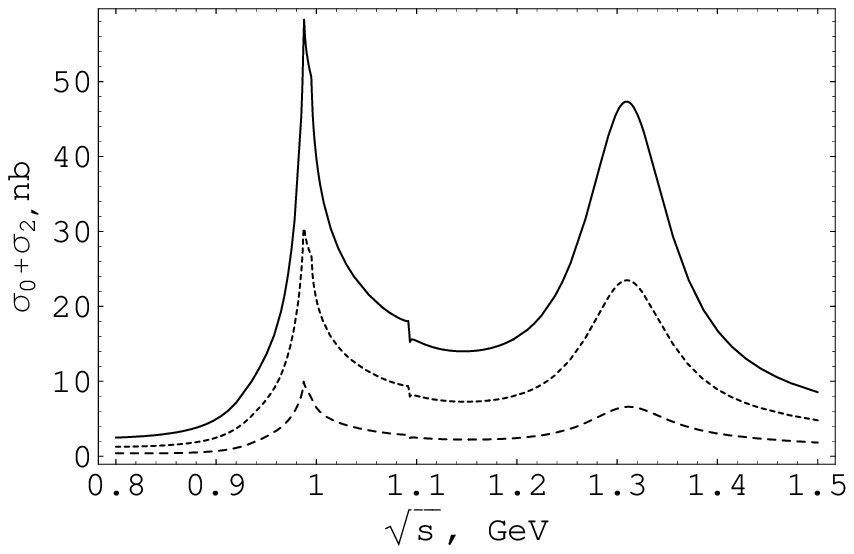}& \includegraphics[width=8cm]{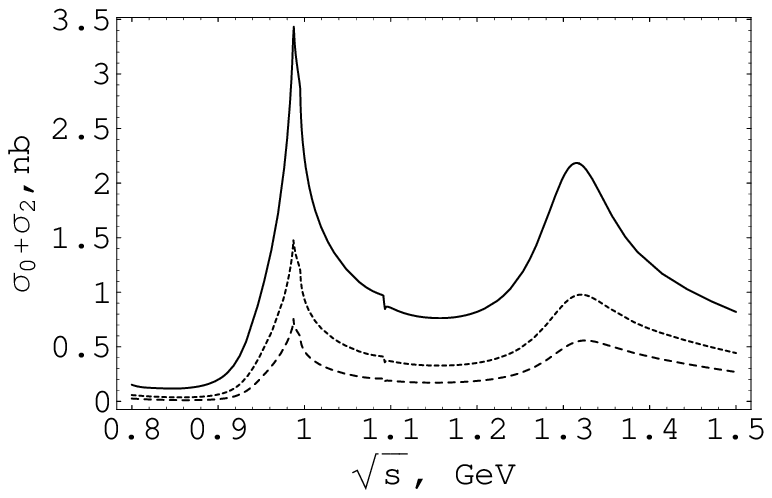}\\ (a)&(b)
\end{tabular}
\end{center}
\caption{The $\sigma_0(\gamma\gamma^*(Q^2)\to \eta\pi^0,
s)+\sigma_2(\gamma\gamma^*(Q^2)\to \eta\pi^0, s)$,
$|\cos\theta|<0.8$, for Fit 1, $a=-0.08$ and $b=-0.15$. (a) Solid
line $Q^2=0$, dashed line $Q^2=0.25$ GeV$^2$, long-dashed line
$Q^2=1$ GeV$^2$; (b) Solid line $Q^2=2.25$ GeV$^2$, dashed line
$Q^2=4$ GeV$^2$, long-dashed line $Q^2=6.25$ GeV$^2$. }
\label{GGcrossQ3}
\end{figure}

\begin{figure}
\begin{center}
\begin{tabular}{ccc}
\includegraphics[width=8cm]{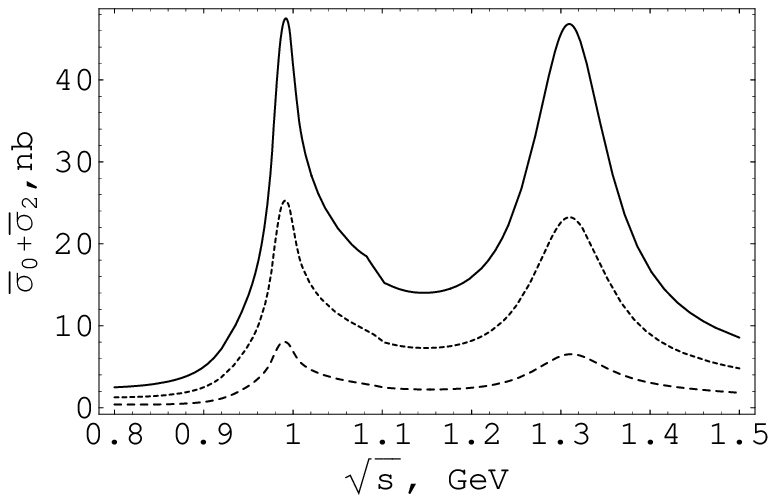}& \includegraphics[width=8cm]{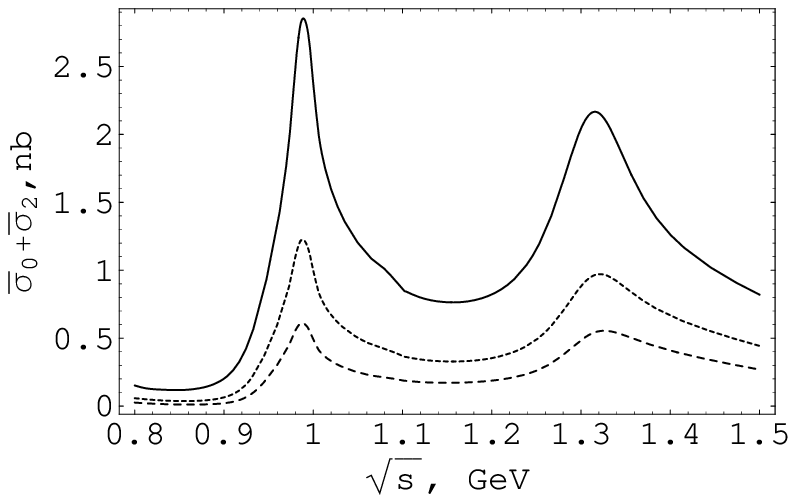}\\ (a)&(b)
\end{tabular}
\end{center}
\caption{The average $\gamma\gamma^*(Q^2)\to \eta\pi^0$ cross
section, $|\cos\theta|<0.8$, for the same parameters as in Fig.
\ref{GGcrossQ3}. Each point of the curve is the cross section
averaged over the $\pm 10$ MeV neighborhood. (a) Solid line
$Q^2=0$, dashed line $Q^2=0.25$ GeV$^2$, long-dashed line $Q^2=1$
GeV$^2$; (b) Solid line $Q^2=2.25$ GeV$^2$, dashed line $Q^2=4$
GeV$^2$, long-dashed line $Q^2=6.25$ GeV$^2$. }
\label{GGcrossQ3Smeared}
\end{figure}

\begin{figure}
\begin{center}
\includegraphics[width=10cm]{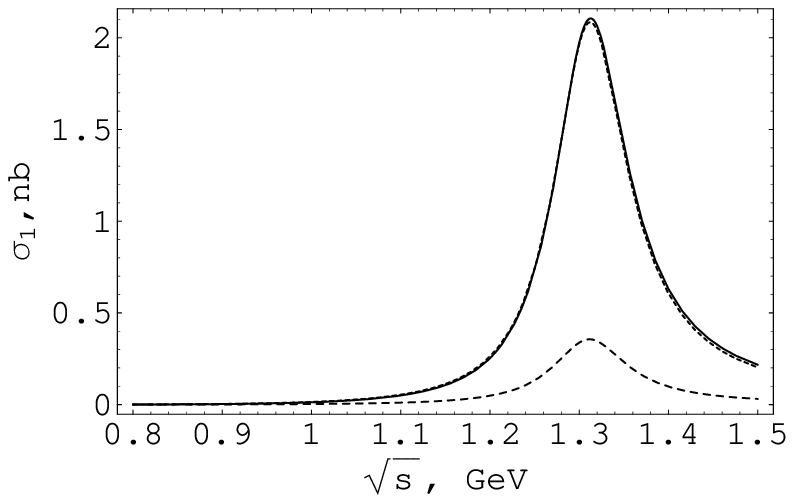}
\end{center}
\caption{The $\sigma_1(\gamma\gamma^*(Q^2)\to \eta\pi^0, s)$,
$|\cos\theta|<0.8$, for the same parameters as in Fig.
\ref{GGcrossQ3}. Solid line $Q^2=0.25$ GeV$^2$, dashed line
$Q^2=1$ GeV$^2$, long-dashed line $Q^2=6.25$ GeV$^2$.}
\label{GGcross1Q3}
\end{figure}

\section{Non-zero Q}
\label{vectorExcitations}

At $Q\to\infty$ the matrix elements in Eqs. (\ref{M0}),
(\ref{M1}), and (\ref{M2}) have the following asymptotics. The
$M^{\mbox{\scriptsize{Born}} \,V}_\lambda$ fall exponentially
because of factor Eq. (\ref{Gfactor}), the kaon loop contribution
$\widetilde{I}^{K^+}_{K^+K^-}(s,Q) T_{K^+K^-\to \pi^0\eta}(s)$ is
$\sim \ln ( Q^2/m^2_{K^+})/Q^2$ \cite{achasov-89prep}. The direct
$a_0\to\gamma^*\gamma$ interaction Eq. (\ref{Mdirect}) does not
depend on $Q$, this is similar to QCD based asymptotics Refs.
\cite{chernyakZhitnitsky-1984,bayerGrozin1985} for a scalar $q\bar
q$ meson transition to $\gamma^*\gamma$. The obtained asymptotics
for the kaon loop contribution $\sim 1/Q^2$ is similar to the QCD
asymptotics for the scalar four-quark meson transition to
$\gamma^*\gamma$. This emphasizes that the direct transition goes
throw the $q\bar q$ component of the $a_0$, and the kaon loop
involves the four-quark component of the $a_0$.

For the $\gamma^*\gamma\to a_2$ transition the obtained
asymptotics are \begin{equation} M_0(\gamma^*\gamma\to a_2)\sim
F_{a_2}(Q) Q^4 , \end{equation}
\begin{equation} M_1(\gamma^*\gamma\to a_2)\sim F_{a_2}(Q) Q^3 , \end{equation}
\begin{equation} M_2(\gamma^*\gamma\to a_2)\sim F_{a_2}(Q) Q^2 . \end{equation}

The hierarchy $M_0:M_1:M_2=Q^2:Q:1$ is the consequence of the
Lagrangian Eq. (\ref{lagra2vv}) and agrees with the QCD based
prediction $M_\lambda(\gamma^*\gamma\to a_2)\sim Q^{-\lambda}$
\cite{chernyakZhitnitsky-1984,bayerGrozin1985}. To reach this
asymptotics one have to conclude that $F_{a_2}(Q) \sim 1/Q^4$,
which is possible when vector excitations $\rho'$, $\rho''$,
$\omega'$, and $\omega''$ are taken into account. Basing on Ref.
\cite{agkr-2013}, let us take
$F_{a_2}(Q)=\widetilde{F}_{a_2}(Q)/\widetilde{F}_{a_2}(0)$, where

$$\widetilde{F}_{a_2}(Q)=\frac{g_{a_2\rho\omega}}{f_\rho
f_\omega}\bigg(\frac{1}{1+Q^2/m_\rho^2}+\frac{1}{1+Q^2/m_\omega^2}\bigg)
+\frac{g_{a_2\rho'\omega'}}{f_{\rho'}
f_{\omega'}}\bigg(\frac{1}{1+Q^2/m_{\rho'}^2}+\frac{1}{1+Q^2/m_{\omega'}^2}\bigg)$$
$$+\frac{g_{a_2\rho''\omega''}}{f_{\rho''}
f_{\omega''}}\bigg(\frac{1}{
1+Q^2/m_{\rho''}^2}+\frac{1}{1+Q^2/m_{\omega''}^2}\bigg)=\frac{g_{a_2\rho\omega}}{f_\rho
f_\omega}\Bigg(\frac{1}{1+Q^2/m_\rho^2}+\frac{1}{1+Q^2/m_\omega^2}$$
\begin{equation}
+a\bigg(\frac{1}{1+Q^2/m_{\rho'}^2}+\frac{1}{1+Q^2/m_{\omega'}^2}\bigg)+b\bigg(\frac{1}{
1+Q^2/m_{\rho''}^2}+\frac{1}{1+Q^2/m_{\omega''}^2}\bigg)\Bigg) .
\label{F2}
\end{equation}

\begin{figure}
\begin{center}
\begin{tabular}{ccc}
\includegraphics[width=8cm]{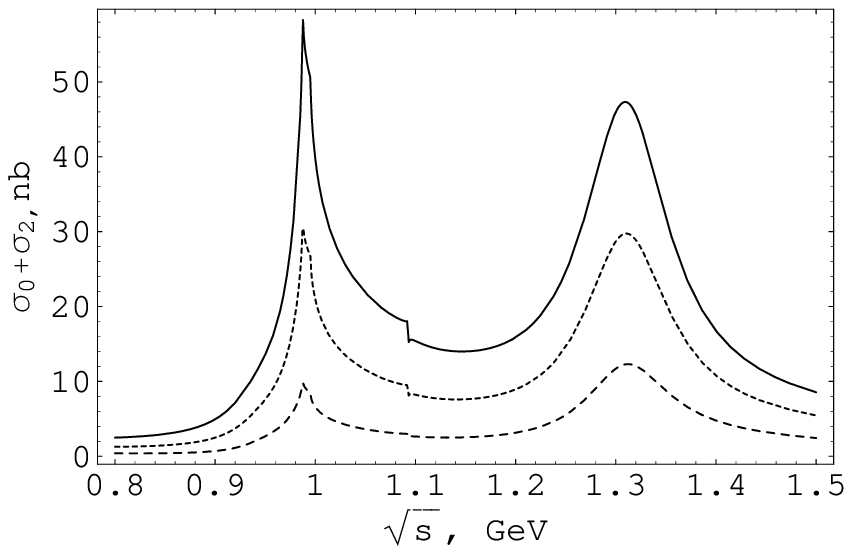}& \includegraphics[width=8cm]{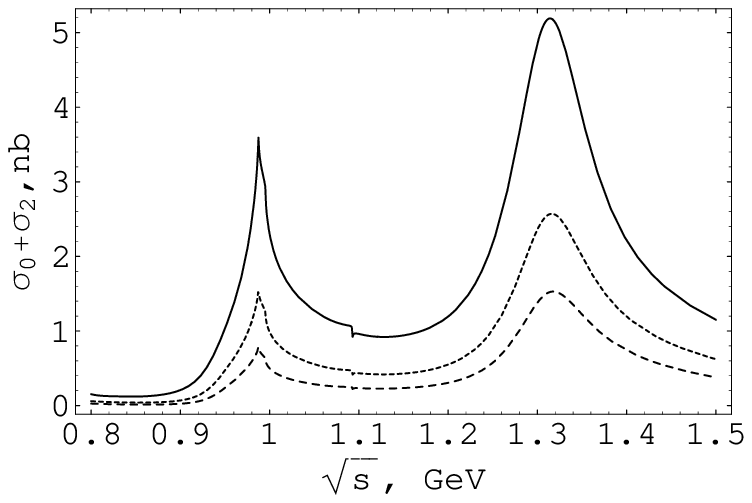}\\ (a)&(b)
\end{tabular}
\end{center}
\caption{The $\sigma_0(\gamma\gamma^*(Q^2)\to \eta\pi^0,
s)+\sigma_2(\gamma\gamma^*(Q^2)\to \eta\pi^0, s)$,
$|\cos\theta|<0.8$, for Fit 1, $a=3.84$ and $b=-3$. (a) Solid line
$Q^2=0$, dashed line $Q^2=0.25$ GeV$^2$, long-dashed line $Q^2=1$
GeV$^2$; (b) Solid line $Q^2=2.25$ GeV$^2$, dashed line $Q^2=4$
GeV$^2$, long-dashed line $Q^2=6.25$ GeV$^2$. } \label{GGcrossQ}
\end{figure}

\begin{figure}
\begin{center}
\includegraphics[width=10cm]{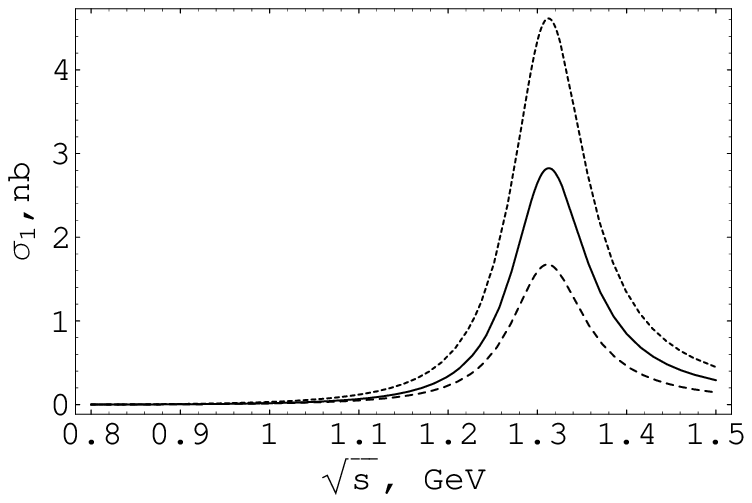}
\end{center}
\caption{The $\sigma_1(\gamma\gamma^*(Q^2)\to \eta\pi^0, s)$,
$|\cos\theta|<0.8$, for the same parameters as in Fig.
\ref{GGcrossQ}. Solid line $Q^2=0.25$ GeV$^2$, dashed line $Q^2=1$
GeV$^2$, long-dashed line $Q^2=6.25$ GeV$^2$.} \label{GGcross1Q}
\end{figure}

\begin{figure}
\begin{center}
\begin{tabular}{ccc}
\includegraphics[width=8cm]{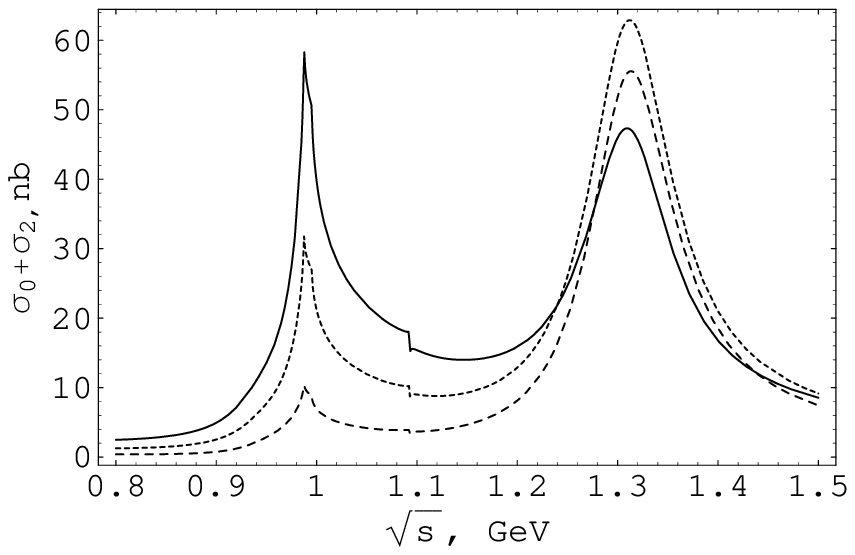}& \includegraphics[width=8cm]{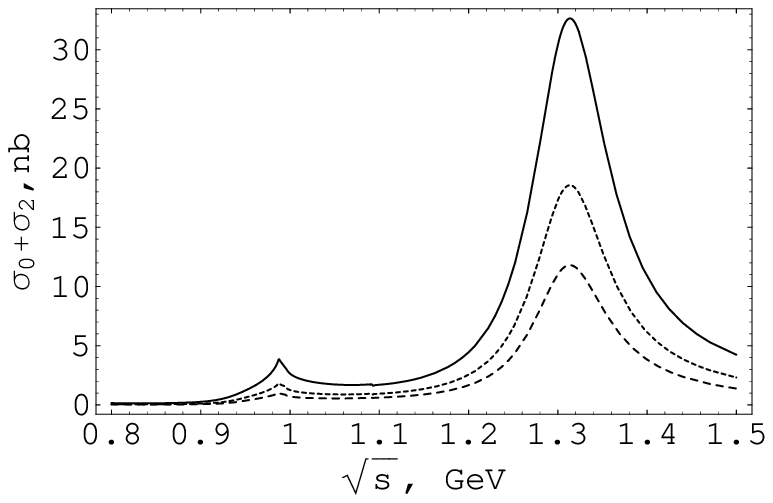}\\ (a)&(b)
\end{tabular}
\end{center}
\caption{The $\sigma_0(\gamma\gamma^*(Q^2)\to \eta\pi^0,
s)+\sigma_2(\gamma\gamma^*(Q^2)\to \eta\pi^0, s)$,
$|\cos\theta|<0.8$, for Fit 1, $a=-4.41$ and $b=3$. (a) Solid line
$Q^2=0$, dashed line $Q^2=0.25$ GeV$^2$, long-dashed line $Q^2=1$
GeV$^2$; (b) Solid line $Q^2=2.25$ GeV$^2$, dashed line $Q^2=4$
GeV$^2$, long-dashed line $Q^2=6.25$ GeV$^2$. } \label{GGcrossQ2}
\end{figure}

\begin{figure}
\begin{center}
\includegraphics[width=10cm]{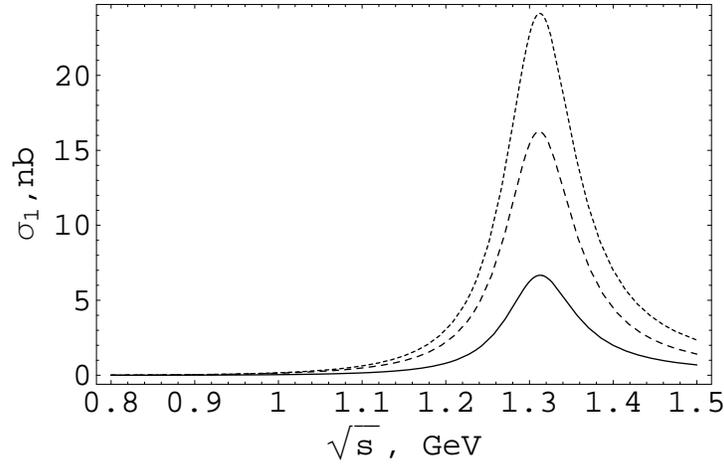}
\end{center}
\caption{The $\sigma_1(\gamma\gamma^*(Q^2)\to \eta\pi^0, s)$,
$|\cos\theta|<0.8$, for the same parameters as in Fig.
\ref{GGcrossQ2}. Solid line $Q^2=0.25$ GeV$^2$, dashed line
$Q^2=1$ GeV$^2$, long-dashed line $Q^2=6.25$ GeV$^2$.}
\label{GGcross1Q2}
\end{figure}

Here $a=g_{a_2\rho'\omega'}f_\rho
f_\omega/g_{a_2\rho\omega}f_{\rho'} f_{\omega'}$ and
$b=g_{a_2\rho''\omega''}f_\rho
f_\omega/g_{a_2\rho\omega}f_{\rho''} f_{\omega''}$. The
requirement

\begin{equation}
m_\rho^2+m_\omega^2+a(m_{\rho'}^2+m_{\omega'}^2)
+b(m_{\rho''}^2+m_{\omega''}^2)=0 \label{zeroing}
\end{equation}

\noindent leads to elimination of the contribution proportional to
$1/Q^2$, so $F_{a_2}(Q)$ is $\sim 1/Q^4$ satisfying the QCD based
asymptotics. This asymptotics is a manifestation of the radial
vector excitations.

The requirement Eq. (\ref{zeroing}) do not fix both $a$ and $b$.
The ratio $a/b$ is the question for the experiment.

In the $\gamma^*(Q^2)\gamma\to f_2(1270)\to\pi^0\pi^0$ reaction
the situation with asymptotics and cancellation due to radial
vector excitations is similar.

Let us take in Eq. (\ref{F2}) $m_{\rho'}=m_{\omega'}=1450$ MeV and
$m_{\rho''}=m_{\omega''}=1700$ MeV in agreement with Ref.
\cite{pdg-2014}.

If one takes $a=-0.08,b=-0.15$ together with the parameters of Fit
1, the $a_0$ and $a_2$ peaks fall synchronously with $Q$ increase,
see Figs. \ref{GGcrossQ3} and \ref{GGcrossQ3Smeared}, where the
sums $\sigma_0 +\sigma_2\equiv \sigma_0(s,Q^2) +\sigma_2(s,Q^2)$
and $\bar \sigma_0 +\bar \sigma_2\equiv \bar \sigma_0(s,Q^2) +\bar
\sigma_2(s,Q^2)$ are shown for different values of $Q^2$. The
$a_2$ peak starts dominating only at $Q\sim 5$ GeV. The $\sigma_1$
is shown for different values of $Q^2$ in Fig. \ref{GGcross1Q3}.
In this case contribution of vector excitations is relatively
small at $Q=0$ as one could expect from general considerations.

Eqs. (\ref{F2}), (\ref{zeroing}) also contain cases when
contribution of vector excitations exceeds the $\rho$ and $\omega$
one even at $Q=0$. Let us consider two of them.

In case $a=3.84$ and $b=-3$ the $a_0$ and $a_2$ peaks in
$\sigma_0+\sigma_2$ fall synchronously for $Q<1$ GeV, then the
$a_2$ peak starts dominating, see Fig. \ref{GGcrossQ}. The
$\sigma_1$ for this case is shown in Fig. \ref{GGcross1Q}.

The choice $a=-4.41$ and $b=3$ leads to Fig. \ref{GGcrossQ2}: the
$a_2$ peak grows at low $Q$ and falls under the value at $Q=0$
only at $Q>1$ GeV. Starting from $Q\approx 0.5$ GeV the $a_2$ peak
dominates over the $a_0$ peak, see Fig. \ref{GGcrossQ2}. The
$\sigma_1$ is shown in Fig. \ref{GGcross1Q2}.

We don't consider scenarios which seem doubtful. For example, for
vast region of $a$ and $b$ the $F_{a_2}(Q_0)=0$ at some $Q_0<1$
GeV.

Note that variation of the parameters in Table I do not provide
such dramatic changes at $Q> 0$ as possible variation of $a$ and
$b$.

New Belle experiment could clarify what scenario is realized.

How vector excitations influence on contributions to the
$\gamma\gamma\to\eta\pi^0$ amplitude not involving the $a_2$ meson
is the question for separate investigation. The kaon loop
contribution probably does not change dramatically because the
couplings of the radial vector excitations with kaon channel,
obtained in \cite{isoscalar}, are small. Other contributions are
not so large. The direct transition $M^{\mbox
{\scriptsize{direct}}}_{\mbox{\scriptsize{res}}}$ is small in
agreement with the four-quark model scenario.

\section{Conclusion}
\label{conclusion}

The experimental data on the $\gamma\gamma\to\eta\pi^0$ reaction
evidence in favor of the four-quark model of the $a_0(980)$. The
data is well described with the scenario based on four-quark
model: relations Eqs. (\ref{FourQrelations}) and small
$g^{(0)}_{a_0\gamma\gamma}$. The obtained values of
$g^{(0)}_{a_0\gamma\gamma}$ mean that the $a_0\phi\gamma$
point-like coupling gives negligible contribution to
$\phi\to\eta\pi^0\gamma$ process. The production (and decay) of
the $a_0(980)$ via rescatterings, i.e. via the four-quark
transitions, is the main qualitative argument in favour of the
four-quark nature of the $a_0(980)$.

The $Q^2$ dependence of the cross section obtained with specific
values of $a$ and $b$ (for example, shown in Figs.
\ref{GGcrossQ3}, \ref{GGcrossQ3Smeared} and \ref{GGcross1Q3}) may
be quite reliable in the region $Q<1$--$2$ GeV \cite{akar-86}. At
higher $Q$ the obtained dependence may be treated only as a guide.
Note that at very high $Q$ one should use QCD.

The strong influence of radial vector excitations at $Q\sim 1$ GeV
is considered also, see Figs. \ref{GGcrossQ}-\ref{GGcross1Q2}.

We don't use the kaon formfactor $G_{K^+}(t,u)$, introduced in
Refs. \cite{annsgn2010,annsgn2011}, since the data can be
explained without it, and the processes $\phi\to\eta\pi^0\gamma$
as well as $\phi\to\pi^0\pi^0\gamma$ are described without this
factor too.

As for comparative production of $a_0$ and $a_2$ in
$\sigma_0+\sigma_2$, at high $Q\gtrsim 5$ GeV the $a_2$
contribution dominates in all variants. In the intermediate region
$Q\sim 1$ GeV the $a_0$ and $a_2$ peaks fall synchronously if
vector excitations contribution is relatively small at $Q=0$ as
one could expect from general considerations. But in principle it
is not excluded that the $a_2$ peak dominates in the intermediate
region too, see Figs. \ref{GGcrossQ}, \ref{GGcrossQ2}.

The data on $\gamma^*(Q^2)\gamma\to \pi^0\pi^0$ have recently
appeared \cite{BellePi0Pi0-2015}. The perspectives of study of the
$f_0(980)$ and $f_2(1270)$ comparative production in this reaction
are poor because the $f_0(980)$ peak is considerably less than the
$f_2(1270)$ peak in $\gamma\gamma\to \pi^0\pi^0$ already
\cite{BellePi0Pi0,annsgn2011}. As it was mentioned above, the
mechanism of $f_2(1270)$ production in the reaction
$\gamma*(Q^2)\gamma\to f_2(1270)$ is similar to the mechanism of
$a_2(1320)$ production in the process $\gamma*(Q^2)\gamma\to
a_2(1320)$. In Ref. \cite{aksh-2015}, we considered in detail the
changeover of the dominant helicity amplitude in the processes
$\gamma*(Q^2)\gamma\to f_2(1270)$ and $\gamma*(Q^2)\gamma\to
a_2(1320)$ with increasing $Q^2$ and showed that data from Ref.
\cite{BellePi0Pi0-2015} could be satisfactorily described even
with only one radial excitation ($\rho'$).

Emphasize that the best process to study the $a_2$ production is
the $\gamma^*(Q^2)\gamma\to a_2\to\rho\pi$ reaction because
$Br(a_2\to\rho\pi)=70\%$ \cite{pdg-2014} and the background is
expected to be small.

The forthcoming SuperKEKB factory, which will have the luminosity
at $40$ times more than the KEKB one \cite{pdg-2014}, could be the
best place for investigation of the scalar and tensor mesons in
$\gamma^*\gamma$ collisions.

\section{Acknowledgements}

This work was supported in part by the Russian Foundation for
Basic Research (Grants Nos. 13- 02-00039 and 16-02-00065) and
Interdisciplinary Project No. 102 of the Siberian Branch, Russian
Academy of Sciences.

\section{Appendix I: Polarization operators of the $a_0$ and $a_0'$}
\label{polarizationOp}

For pseudoscalar $a,b$ mesons and $m_a\geq m_b,\ m\geq m_+$ one
has:

\begin{eqnarray}
\label{polarisator}
&&\Pi^{ab}_R(m^2)=\frac{g^2_{Rab}}{16\pi}\left[\frac{m_+m_-}{\pi
m^2}\ln \frac{m_b}{m_a}+\right.\nonumber\\
&&\left.+\rho_{ab}\left(i+\frac{1}{\pi}\ln\frac{\sqrt{m^2-m_-^2}-
\sqrt{m^2-m_+^2}}{\sqrt{m^2-m_-^2}+\sqrt{m^2-m_+^2}}\right)\right]
.
\end{eqnarray}
\noindent For $m_-\leq m<m_+$
\begin{eqnarray}
&&\Pi^{ab}_{R}(m^2)=\frac{g^2_{Rab}}{16\pi}\left[\frac{m_+m_-}{\pi
m^2}\ln \frac{m_b}{m_a}-|\rho_{ab}(m)|+\right.\nonumber\\
&&\left.+\frac{2}{\pi}|\rho_{ab}(m)
|\arctan\frac{\sqrt{m_+^2-m^2}}{\sqrt{m^2-m_-^2}}\right],
\end{eqnarray}
\noindent for $m<m_-$
\begin{eqnarray}
&&\Pi^{ab}_{R}(m^2)=\frac{g^2_{Rab}}{16\pi}\left[\frac{m_+m_-}{\pi
m^2}\ln \frac{m_b}{m_a}-\right.\nonumber\\
&&\left.-\frac{1}{\pi}\rho_{ab}(m)\ln\frac{\sqrt{m_+^2-m^2}-
\sqrt{m_-^2-m^2}}{\sqrt{m_+^2-m^2}+\sqrt{m_-^2-m^2}}\right].
\end{eqnarray}

Note that we take into account intermediate states
$\eta\pi^0,K\bar K,\eta'\pi^0$ in the $a_0(980)$ and $a_0'$
propagators:

\begin{equation}
\Pi_{a_0}=\Pi_{a_0}^{\eta\pi^0}+\Pi_{a_0}^{K^+K^-}+\Pi_{a_0}^{K^0\bar
K^0}+ \Pi_{a_0}^{\eta'\pi^0} ,
\end{equation}

\noindent and the same for $a_0'$. Note that
$g_{a_0K^+K^-}=-g_{a_0K^0\bar{K^0}}$ and
$g_{a_0'K^+K^-}=-g_{a_0'K^0\bar{K^0}}$.

\section{Appendix II: The $\eta\pi^0$ spectrum in $\phi\to\eta\pi^0\gamma$ decay}
\label{phiSpectr}

The amplitude of the signal process $\phi(p)\to \gamma(a_0+a_0')
\to\gamma(q)\pi^0(k_1)\eta(k_2)$ is

\begin{equation}
M_{sig}=e^{i\delta_B}g(m)\sum_{R,R'}g_{RK^+K^-}G_{RR'}^{-1}g_{R'\eta\pi^0}\,\bigg(
(\phi\epsilon)- \frac{(\phi q)(\epsilon p)}{(pq)} \bigg)
\label{a0signal}\,,
\end{equation}
where  $m^2=(k_1+k_2)^2$, $\phi_{\alpha}$ and $\epsilon_{\mu}$ are
the polarization vectors of $\phi$ meson and photon, the function
$g(m)$ is given below. The
$\delta_B=\delta^{bg}_{\eta\pi^0}+\delta^{bg}_{K\bar K}$.

The matrix element of the background process
$\phi(p)\to\pi^0\rho^0\to\gamma(q)\pi^0(k_1)\eta(k_2)$ is

\begin{equation}
M_B=\frac{g_{\phi\rho\pi}g_{\rho\eta\gamma}}{D_{\rho}(p-k_1)}
\phi_{\alpha}k_{1\mu}p_{\nu}\epsilon_{\delta}(p-k_1)_{\omega}q_{\epsilon}
\epsilon_{\alpha\beta\mu\nu}\epsilon_{\beta\delta\omega\epsilon}.\label{Mback}
\end{equation}

The mass spectrum is
\begin{equation}
\frac{d\Gamma(\phi\to\gamma\pi^0\eta,m)}{dm}=\frac{d\Gamma_{sig}(m)}{dm}+
\frac{d\Gamma_{back}(m)}{dm}+ \frac{d\Gamma_{int}(m)}{dm}\,,
\end{equation}
where the mass spectrum for the signal is
\begin{equation}\frac{d\Gamma_{sig}(m)}{dm}=
\frac{2|g(m)|^2p_{\eta\pi}(m_{\phi}^2-m^2)}
{3(4\pi)^3m_{\phi}^3}\bigg|\sum_{R,R'}g_{RK^+K^-}G_{RR'}^{-1}g_{R'\eta\pi^0}\bigg|^2\,.
\label{spectruma0}
\end{equation}

The mass spectrum for the background process
$\phi\to\pi^0\rho\to\gamma\pi^0\eta$ is \cite{a0f0}:

\begin{equation}
\frac{d\Gamma_{back}(m)}{dm}=\frac{(m_{\phi}^2-m^2)p_{\pi\eta}
}{128\pi^3m_{\phi}^3}\int_{-1}^{1}dxA_{back}(m,x)\,,
\end{equation}
where
\begin{eqnarray}
&&A_{back}(m,x)=\frac{1}{3}\sum|M_B|^2= \nonumber \\
&&=\frac{1}{24}(m_{\eta}^4m_{\pi}^4+2m^2m_{\eta}^2m_{\pi}^2
\tilde{m_{\rho}}^2-2m_{\eta}^4m_{\pi}^2\tilde{m_{\rho}}^2-2m_{\eta}^2m_{\pi}^4
\tilde{m_{\rho}}^2+\nonumber \\ &&2m^4\tilde{m_{\rho}}^4-
2m^2m_{\eta}^2\tilde{m_{\rho}}^4+m_{\eta}^4\tilde{m_{\rho}}^4
-2m^2m_{\pi}^2\tilde{m_{\rho}}^4+4m_{\eta}^2m_{\pi}^2\tilde{m_{\rho}}^4
+m_{\pi}^4\tilde{m_{\rho}}^4+\nonumber \\
&&2m^2\tilde{m_{\rho}}^6-
2m_{\eta}^2\tilde{m_{\rho}}^6-2m_{\pi}^2\tilde{m_{\rho}}^6+
\tilde{m_{\rho}}^8-2m_{\eta}^4m_{\pi}^2m_{\phi}^2-
2m^2m_{\eta}^2m_{\phi}^2\tilde{m_{\rho}}^2+\nonumber \\
&&2m_{\eta}^2m_{\pi}^2m_{\phi}^2\tilde{m_{\rho}}^2-
2m^2m_{\phi}^2\tilde{m_{\rho}}^4+
2m_{\eta}^2m_{\phi}^2\tilde{m_{\rho}}^4-
2m_{\phi}^2\tilde{m_{\rho}}^6+ m_{\eta}^4m_{\phi}^4+
m_{\phi}^4\tilde{m_{\rho}}^4)\times\nonumber \\
&&\bigg|\frac{g_{\phi\rho\pi}g_{\rho\eta\gamma}}
{D_{\rho}(\tilde{m_{\rho}})}\bigg|^2\,, \label{Aback}
\end{eqnarray}
and
\begin{eqnarray}
&&\tilde{m_{\rho}}^2=m_{\eta}^2+\frac{(m^2+m_{\eta}^2-m_{\pi}^2)(m_{\phi}^2-
m^2)}{2m^2}-\frac{(m_{\phi}^2-m^2)x}{m}p_{\pi\eta}\nonumber \\
&&p_{\pi\eta}=\frac{\sqrt{(m^2-(m_{\eta}-m_{\pi})^2)
(m^2-(m_{\eta}+m_{\pi})^2)}}{2m}\,.
\end{eqnarray}

The term of the interference between the signal and the background
processes is written in the following way:

\begin{equation}
\frac{d\Gamma_{int}(m)}{dm}=\frac{(m_{\phi}^2-m^2)p_{\pi\eta}
}{128\pi^3m_{\phi}^3} \int_{-1}^{1}dxA_{int}(m,x)\,,
\end{equation}
where
\begin{eqnarray}
&&A_{int}(m,x)=\frac{2}{3}Re\sum M_{sig}M_B^*=
\frac{1}{3}\left((m^2-m_{\phi}^2)\tilde{m_{\rho}}^2+
\frac{m_{\phi}^2(\tilde{m_{\rho}}^2-m_{\eta}^2)^2}{m_{\phi}^2-m^2}\right)
\times\nonumber\\ &&Re\bigg\{\frac{e^{i\delta
}g(m)\Big(\sum_{R,R'}g_{RK^+K^-}G_{RR'}^{-1}g_{R'\eta\pi^0}\Big)g_{\phi\rho\pi}g_{\rho\eta\gamma}}
{D^*_{\rho}(\tilde{m_{\rho}})}\bigg\}\,.
\end{eqnarray}

The $\delta$ is additional relative phase between $M_{sig}$ and
$M_B$, does not mentioned in Eqs. (\ref{a0signal}) and
(\ref{Mback}). It is assumed to be constant and takes into
account, for example, $\rho\pi$ rescattering effects, see
\cite{rhophase}.

In the $\phi\to K^+K^-\to a_0\gamma$ loop model $g(m)$ has the
following forms:

\noindent for $m<2m_{K^+}$

\begin{eqnarray}
&&g(m)=\frac{e}{2(2\pi)^2}g_{\phi K^+K^-}\Biggl\{
1+\frac{1-\rho_{K^+K^-}^2(m^2)}{\rho_{K^+K^-}^2(m^2_{\phi})-\rho_{K^+K^-}^2(m^2)}\times\nonumber\\
&&\Biggl[2|\rho_{K^+K^-}(m^2)|\arctan\frac{1}{|\rho_{K^+K^-}(m^2)|}
-\rho_{K^+K^-}(m^2_{\phi})\lambda(m^2_{\phi})+i\pi\rho_{K^+K^-}(m^2_{\phi})-\nonumber\\
&&-(1-\rho_{K^+K^-}^2(m^2_{\phi}))\Biggl(\frac{1}{4}(\pi+
i\lambda(m^2_{\phi}))^2- \nonumber\\
&&-\Biggl(\arctan\frac{1}{|\rho_{K^+K^-}(m^2)|}\Biggr)^2
\Biggr)\Biggr]\Biggr\},
\end{eqnarray}
where
\begin{equation}
\lambda(m^2)=\ln\frac{1+\rho_{K^+K^-}(m^2)}{1-\rho_{K^+K^-}(m^2)}\,\,;\qquad
\frac{e^2}{4\pi}=\alpha=\frac{1}{137}\,\,.
\end{equation}

For $m\geq 2m_{K^+}$
\begin{eqnarray}
&&g(m)=\frac{e}{2(2\pi)^2}g_{\phi K^+K^-}\Biggl\{
1+\frac{1-\rho_{K^+K^-}^2(m^2)}{\rho_{K^+K^-}^2(m^2_{\phi})-\rho_{K^+K^-}^2(m^2)}\times\nonumber\\
&&\times\Biggl[\rho_{K^+K^-}(m^2)(\lambda(m^2)-i\pi)-
\rho_{K^+K^-}(m^2_{\phi})(\lambda(m^2_{\phi})-i\pi)-\nonumber\\
&&\frac{1}{4}(1-\rho_{K^+K^-}^2(m^2_{\phi}))
\Biggl((\pi+i\lambda(m^2_{\phi}))^2-
(\pi+i\lambda(m^2))^2\Biggr)\Biggr]\Biggr\}.
\end{eqnarray}

Note that $g(m)\to -\widetilde{I}^{K^+}_{K^+K^-}(s,Q)/16\pi e$,
see Eq. (\ref{IKK}), for $m^2\to s,\,m_\phi^2\to -Q^2$.

The inverse propagator of the $\rho$ meson has the following
expression
\begin{equation}
D_{\rho}(m)=m_{\rho}^2-m^2-im^2\frac{g^2_{\rho\pi\pi}}{48\pi}
\bigg(1-\frac{4m_{\pi}^2}{m^2}\bigg)^{3/2}\,.
\end{equation}

We use coupling constants $g_{\phi K^+K^-}=4.47$,
$g_{\phi\rho\pi}=0.803$ GeV$^{-1}$ and $g_{\rho\eta\gamma}=0.50$
GeV$^{-1}$, obtained with the help of Ref. \cite{pdg-2014} data.

\section{Appendix III: Additional terms in $\chi^2$ function}
\label{chi2AddSection}

The function to minimize, $\chi^2$ function, in the present paper
consists of three terms:

\begin{equation}
\chi^2=\chi^2_{\gamma\gamma}+\chi^2_{spec}+\chi^2_{add}.
\end{equation}

The $\chi^2_{\gamma\gamma}$ and $\chi^2_{spec}$ are usual $\chi^2$
functions for the $\sigma(\gamma\gamma\to\eta\pi^0,s)$ and
$\eta\pi^0$ spectrum of the $\phi\to\eta\pi^0\gamma$ reaction.

The $\chi^2_{add}$ represents additional terms to reach some
restrictions: lower $f_{K\bar K}$ to reduce influence of the
analytical continuation of the $e^{i\delta^{bg}_{K\bar K}}$ under
the $K\bar K$ threshold on the $|M_{sig}|$, see Eq.
(\ref{a0signal}); coupling constants close to the relations Eq.
(\ref{FourQrelations}); not large $g^{(0)}_{a_0 \gamma\gamma}$ and
not very large $f_{\pi\eta'}$. The $\chi^2_{add}$ was used to
obtain Fits 1, 3, and 4. Fit 2 without $\chi^2_{add}$ shows that
the $a_0$ parameters not go far from the Fit 1 results.


\begin{thebibliography}{99}
\bibitem{a0prime}
One can see from Table I that the $a_0'$ parameters are not
determined well on the present stage. The nature of this particle
currently can not be fixed. For example, it is possible that
$a_0'$ is a complex of several scalars, effectively taken into
account. The question of the $a_0'$ parameters and nature is not
important for the aims of the present work.
\bibitem{achasov-89prep}
N.N. Achasov, Phys. Lett. B {\bf 222}, 139 (1989). It results from
Eq. (7) of this paper, as for the last line of Eq. (12), it has a
typo.
\end{thebibliography}
\end{document}